\documentclass[a4paper,11pt]{article}
\begin{document}
\def\boxit#1{\vcenter{\hrule\hbox{\vrule\kern8pt
      \vbox{\kern8pt#1\kern8pt}\kern8pt\vrule}\hrule}}
\def\Boxed#1{\boxit{\hbox{$\displaystyle{#1}$}}} 
\def\sqr#1#2{{\vcenter{\vbox{\hrule height.#2pt
        \hbox{\vrule width.#2pt height#1pt \kern#1pt
          \vrule width.#2pt}
        \hrule height.#2pt}}}}
\def\square{\mathchoice\sqr34\sqr34\sqr{2.1}3\sqr{1.5}3}
\def\Square{\mathchoice\sqr67\sqr67\sqr{5.1}3\sqr{1.5}3}
\def\AJP{{\it Am. J. Phys.}}
\def\AM{{\it Ann. Math.}}
\def\AP{{\it Ann. Phys.}}
\def\CQG{{\it Class. Quant. Grav.}}
\def\GRG{{\it Gen. Rel. Grav.}}
\def\JMP{{\it J. Math. Phys.}}
\def\JP{{\it J. Phys.}}
\def\JSIRAN{{\it J. Sci. I. R. Iran}}
\def\NC{{\it Nuovo Cim.}}
\def\NP{{\it Nucl. Phys.}}
\def\PL{{\it Phys. Lett.}}
\def\PR{{\it Phys. Rev.}}
\def\PRL{{\it Phys. Rev. Lett.}}
\def\PRp{{\it Phys. Rep.}}
\def\RMP{{\it Rev. Mod. Phys.}}
%
\title{\bf Third Order Lagrangians, Weyl Invariants \&
           Classical Trace Anomaly in Six Dimensions}
\author{Mehrdad Farhoudi\thanks{E-mail:
 m-farhoudi@sbu.ac.ir}\\
 {\small Department of Physics, Shahid Beheshti University G.C.,
         Evin, Tehran 19839, Iran}}
\date{\small April 30, 2013}
\maketitle
\begin{abstract}
\noindent
 We have proceeded the analogy (represented in our previous works)
of the Einstein tensor and the alternative form of the Einstein
field equations for the generic coefficients of the eight terms in
the third order of the Lovelock Lagrangian. We have found the
constraint between the coefficients into two forms, an independent
and a dimensional dependent versions. Each form has three degrees
of freedom, and not only the exact coefficients of the third order
Lovelock Lagrangian do satisfy the two forms of the constraints,
but also the two independent cubic of the Weyl tensor satisfy the
independent constraint in six dimensions and yield the dimensional
dependent version identically independent of the dimension. Then,
we have introduced the most general effective expression for a
total third order type Lagrangian with the homogeneity degree
number three which includes the previous eight terms plus the new
three ones among the all seventeen independent terms. We have
proceeded the analogy for this combination, and have achieved the
relevant constraint. We have shown that the expressions given in
the literature as the third Weyl--invariant combination in six
dimensions do satisfy this constraint. Thus, we suggest that these
constraint relations to be considered as the necessary consistency
conditions on the numerical coefficients that a Weyl--invariant in
six dimensions should satisfy. Finally, we have calculated the
``classical'' trace anomaly (an approach that was presented in our
previous works) for the introduced total third order type
Lagrangian and have achieved a general expression with four
degrees of freedom in more than six dimensions (three degrees in
six dimensions). Then, we have demonstrated that the resulted
expression contains exactly the relevant coefficient of the
Schwinger--DeWitt proper time method (that linked with the
relevant heat kernel coefficient) in six dimensions, as a
particular case. Of course, this result is a necessary consistency
check, nevertheless our approach can be regarded as an alternative
(perhaps simpler, and classical) derivation of the trace anomaly
which also gives a general expression with the relevant degrees of
freedom.
\end{abstract}
\bigskip
\section{Introduction}
\indent

In our previous work~\cite{farb}, we have highlighted that the
splitting feature of the Einstein tensor (as the first term of the
Lovelock tensor~\cite{lovedcbrig}) into two parts -- namely the
Ricci tensor and the term proportional to the curvature scalar --
with the trace relation between them is a common feature of each
homogeneous term in the Lovelock tensor. We have emphasized that,
indeed, this property can been resulted through the variation
procedure. Then, motivated by the principle of general covariance,
we have shown that this property can be generalized, via a
generalized trace operator which we have defined and denoted by
the {\it Trace} notation (instead of the trace one), for any
inhomogeneous Euler--Lagrange expression that can be spanned
linearly in terms of homogeneous tensors. And, as an immediate
application, we have demonstrated that the (whole) Lovelock tensor
(which is constructed of terms with a mixture of different orders)
analogizes the mathematical feature of the Einstein tensor, and
hence, we have classified the Lovelock gravity as a generalized
Einstein gravity.

Still motivated by the principle of general covariance, we have
proceeded~\cite{farc} further the analogy and enforced the
mathematical appearance of the alternative form of the Einstein
field equations (as a covariant form for gravitational field
equations) for the relevant alternative form of the Lovelock field
equations. From this, we have found that the price for this
analogy is to accept the existence of the trace anomaly of the
energy--momentum tensor even in the classical treatments. Thus, we
have actually denoted~\cite{farc} a classical view of gravitation
which explicitly shows the presence of an extra anomalous trace
for the energy--momentum tensor with an indication of the
constitution of the higher order gravities towards it, exactly as
what has been verified~\cite{dufc,deser96,asgosh} in the quantum
aspects of gravity. Indeed, in the quantum theory, one relates and
justifies the presence of the trace anomaly, classically, to
higher order gravities~\cite{dufb}. That is, the higher--order
terms in curvature invariants have to be added to the effective
Lagrangian of gravitational field when quantum corrections are
considered, see, e.g., Ref.~\cite{bida} and the beginning of
Sect.~2. It has also been realized that such corrective terms are
inescapable in order to obtain the effective quantum gravity
action near the Plank scale~\cite{vilko}. Nevertheless, it is
possible to show that, via conformal transformations, the
higher--order (and even non--minimally coupled) terms always
correspond to the Einstein gravity plus one (or more than one)
minimally coupled scalar field, see, e.g.,
Refs.~\cite{fare,CapoLaur} and references therein. Also, We have
interpreted~\cite{fare} that our classical procedure indicates
compatibility with the Mach idea.

Then, we have employed~\cite{farc} this procedure for any generic
coefficients of the second order term of the Lovelock Lagrangian.
And thus, we have gotten the resulted trace anomaly relation
exactly the same as the trace anomaly consistency constraint which
had been suggested by Duff~\cite{dufa} in any dimension. In
addition to the second term of the Lovelock Lagrangian, the action
constructed by the square of the Weyl conformal tensor, which is
the only local geometrical conformal invariant in four dimensions,
also satisfies this constraint\footnote{According to the
classification of Refs.~\cite{deser96,desch}, based on the
dimensional regularization and power counting, this constraint
indicates that in four dimensions (and indeed, in even
dimensions), the anomaly can have two contributions, a type A
anomaly (the Euler density invariants) and a type B anomaly (built
from the conformal invariants). Also, a purely algebraic
classification (independent of any regularization scheme) of the
structure of the Weyl anomalies in arbitrary space--time
dimensions has been presented in Ref.~\cite{Boulanger07}.}\
 relation (only) in four dimensions (note that, the Weyl tensor is
itself identically zero in three dimensions). This exact equality,
between the trace anomaly relation suggested by Duff and the
constraint relation that the constant coefficients of any generic
second order Lagrangian must satisfy in order to hold the desired
analogy, indicates that there may exist an intriguing interplay
here. Indeed, one can speculate that an intrinsic reason for the
existence of such a relation should perhaps be, classically,
somehow related to the covariance\footnote{In the semi--classical
approach of quantum gravity theory, which has been employed to
deduce the trace anomalies, the conformal invariance is
sacrificed~\cite{frw} to the needs of the general covariance
(though, in contrary, see Ref.~\cite{quiros00}).}\
 of the form of Einstein's
equations~\cite{farc}. That is, by employing the generalized trace
relation, the appearance of the trace anomaly may be interpreted
as the Lovelock modification of gravity even in the classical
treatments. Though it is somehow a naive conjecture, nevertheless,
it gives almost an easy classical procedure to grasp the desired
result which, besides what have been already given in the
literature~\cite{Boulanger07,quiros00,bpb,bocrnoodog}, indicates
an intrinsic property behind it. Hence, as a main advantage of
this procedure, we have suggested~\cite{fare} that this analogous
of the Einstein tensor can even be employed as a criteria in order
to distinguish correct/legitimate metric theories of gravity which
are either homogeneous functions, or linear combinations of
different homogeneous functions, of the metric and its
derivatives.

The analogy is also capable to provide dimensional dependent
relations. Actually, we have derived a dimensional dependent
version of the Duff trace anomaly relation in Ref.~\cite{fare}, in
where we have also summarized the important achievements of the
procedure. The Gauss--Bonnet and the Weyl squared terms do satisfy
this dimensional dependent constraint relation in any dimension.
This dimensional dependent version of the Duff trace anomaly
relation has also been re--derived in Ref.~\cite{Oliva2010} (their
relation~(7)) by classifying higher derivative theories of gravity
whose the traced field equations have a reduced order.

Now, in this work, we propose to probe further the analogy for the
field tensor of the corresponding generic coefficients of the
third order term of the Lovelock Lagrangian, and then for any type
of third order Lagrangian. For this purpose, we give a brief
review on the trace idea, particularly on the definition of the
generalized trace tool in the following section. In Sect.~$3$, we
provide the necessary relations for the generic third order
Lagrangian terms and their Euler--Lagrange expressions. Also, in
addition to the eight linearly independent terms, which appear in
the third order of the Lovelock Lagrangian, we specify the other
linearly independent types of third order Lagrangian terms. Then,
in Sect.~$4$, based on the desired analogy, we derive the
constraint relations among the constituent coefficients of the
generic third order Lagrangians in two forms, an independent and a
dimensional dependent versions. Hence, we match these relations
for the two linearly independent scalars formed by the cubic of
the Weyl tensor. The other Weyl--invariants are investigated when
we incorporate all types of third order Lagrangian terms in
Sect.~$5$. Actually, we first probe the trace analogy for the most
general effective expression that we will introduce as a total
third order Lagrangian, and then examine the resulted constraint
relations for the numerical coefficients of the Weyl--invariants.
We indicate the classical view of the trace anomaly in six
dimensions in Sect.~$6$, in where we also demonstrate that it has
an interesting similarity with the appropriate heat kernel
coefficients. Conclusions are presented in the last section, and
some necessary formulations for the metric variation of a few
Lagrangian terms plus a few useful relations and some derivations
are furnished in Appendixes A and B.

\section{The Trace Idea Review}
\indent

We follow the sign conventions of Wald~\cite{wald} and, analogous
to the Einstein field equations, we assume that the geometry is
proportional to the matter in a way that the gravitational field
equations should be $G^{(\rm gravi.)}_{\alpha\beta}=\kappa^2\,
T_{\alpha\beta}/2$, where the gravitational tensor, $G^{(\rm
gravi.)}_{\alpha\beta}$, represents the geometry, $\kappa^2\equiv
16\pi G/c^4$ is the constant of the proportionality and the lower
case Greek indices run from zero to $D-1$ in a $D$--dimensional
space--time. Actually, the gravitational tensor is the
Euler--Lagrange expression given in the process of the metric
variation of the action, $\int L^{(\rm gravi.)}\sqrt{-g}\,
d^D\!x$, i.e.
\begin{equation}\label{defggrav}
\delta\left[L^{(\rm gravi.)}\sqrt{-g}\,\right]=\left[\frac{\delta
L^{(\rm gravi.)}}{\delta
g^{\alpha\beta}}-\frac{1}{2}g_{\alpha\beta}L^{(\rm
gravi.)}\right]\sqrt{-g}\, \delta
g^{\alpha\beta}=\frac{1}{\kappa^2}G^{(\rm gravi.)}_{\alpha\beta}\,
\sqrt{-g}\, \delta g^{\alpha\beta}.
\end{equation}
Also, analogous to the Einstein tensor, we demand
\begin{equation}\label{drFg}
G^{(\rm gravi.)}_{\alpha\beta}=R^{(\rm
gravi.)}_{\alpha\beta}-\frac{1}{2}g_{\alpha\beta}R^{(\rm gravi.)},
\end{equation}
with the trace relation between its two parts, exactly the same as
the one yields between the Ricci tensor and the Ricci scalar.

In addition to the Einstein tensor, the Lovelock tensor has also
been employed as a gravitational tensor. For example, the
superstring theory, in its low energy limit, suggests that the
Einstein--Hilbert action should be enlarged by the inclusion of
higher order curvature terms, and in order to be ghost--free it
has been shown~\cite{zwizum} that it must be in the form of
dimensionally continued Gauss--Bonnet densities (i.e., the
Lovelock Lagrangian terms). An important aspect of this suggestion
is~\cite{ishishb} that it does not arise in attempts to quantize
gravity. The ghost--free property and the fact that the Lovelock
Lagrangian is the most general Lagrangian which (the same as the
Einstein--Hilbert Lagrangian) yields the field equations as second
order equations, have stimulated interests in the Lovelock gravity
and its applications in the literature\rlap.\footnote{For review
on the inclusion of higher order Lagrangians see, e.g.,
Refs.~\cite{fare,CapoLaur,SchCapoSotiNoOdsCF2011CEtAl} and
references therein.}\
 The Lovelock Lagrangian is~\cite{lovedcbrig}
\begin{equation}\label{love}
L^{\rm
(Lovelock)}=\frac{1}{\kappa^2}\sum_{0<n<\frac{D}{2}}\,\frac{1}{2^n}\,
c_n\,\delta^{\alpha_1\ldots\alpha_{2n}}_{\beta_1\ldots\beta_{2n}}\,
R_{\alpha_1\alpha_2}{}^{\beta_1\beta_2}\cdots R_{\alpha_{2n-1}\,
\alpha_{2n}}{}^{\beta_{2n-1}\,\beta_{2n}}
\equiv\sum_{0<n<\frac{D}{2}}\, c_n\, L^{(n)},
\end{equation}
where we set $c_1\equiv 1$ and the other $c_n$ constants are of
the order of the Planck length, $\ell_P=\sqrt{\hbar G/c^3}$, to
the power $2(n-1)$, for making the dimension of $L^{\rm
(Lovelock)}$ to be the same as $L^{(1)}\equiv L_{_{\rm
E-H}}=R/\kappa^2$. The
$\delta^{\alpha_1\ldots\alpha_p}_{\beta_1\ldots\beta_p}$ is the
generalized Kronecker delta symbol, which is identically zero if
$p>D$, and in relation (\ref{love}), the extremum value of $n$ is
related to the dimension of space--time by
\begin{equation}\label{nlim}
 n_{_{\rm ext.}}\!= \cases{
                          \frac{D}{2}-1 & \textrm{even $D$}\cr
                          \cr
                          \frac{D-1}{2} & \textrm{odd {} $D$.}\cr}
\end{equation}
Hence, the $L^{\rm (Lovelock)}$ reduces to the Einstein--Hilbert
Lagrangian in four dimensions, and its second term is the
Gauss--Bonnet invariant. The Lovelock tensor, as dimensionally
reduction Euler--Lagrange terms, is~\cite{lovedcbrig}
\begin{equation}\label{lovet}
G^{\rm (Lovelock)}_{\alpha\beta}\!=-\!\!\!\!\sum_{0<n<\frac{D}{2}}
\!\!\frac{1}{2^{n+1}}\, c_n\, g_{\alpha\mu}\,
\delta^{\mu\alpha_1\ldots\alpha_{2n}}_{\beta\beta_1\ldots\beta_{2n}}
\, R_{\alpha_1\alpha_2}{}^{\beta_1\beta_2}\cdots
R_{\alpha_{2n-1}\,\alpha_{2n}}{}^{\beta_{2n-1}\,\beta_{2n}}\equiv
\!\!\!\!\sum_{0<n<\frac{D}{2}}\!\! c_n\, G^{(n)}_{\alpha\beta}\, ,
\end{equation}
where the cosmological term has been neglected and the
$G^{(1)}_{\alpha\beta}\equiv G_{\alpha\beta}$, i.e. the Einstein
tensor. Relation (\ref{lovet}) can also be written as $G^{(\rm
Lovelock)}_{\alpha\beta}=R^{(\rm
Lovelock)}_{\alpha\beta}-g_{\alpha\beta}R^{(\rm Lovelock)}/2$,
where~\cite{farb}
\begin{equation}\label{rrlove}
R^{\rm (Lovelock)}_{\alpha\beta}\!\equiv
\!\!\!\!\sum_{0<n<\frac{D}{2}}\!\! c_n\,
R^{(n)}_{\alpha\beta}\qquad\qquad {\rm and}\qquad\qquad R^{\rm
(Lovelock)}\!=\kappa^2 L^{\rm (Lovelock)}\equiv
\!\!\!\!\sum_{0<n<\frac{D}{2}}\!\! c_n\, R^{(n)},
\end{equation}
with $R^{(n)}_{\alpha\beta}$ defined as
\begin{equation}\label{ricn}
R^{(n)}_{\alpha\beta}\equiv {n\over 2^n}\,\delta^{\alpha_1\alpha_2%
\ldots\alpha_{2n}}_{\alpha\ \beta_2\ldots\beta_{2n}}\,
R_{\alpha_1\alpha_2\beta}{}^{\beta_2}\, R_{\alpha_3\alpha_4}
{}^{\beta_3\beta_4}\cdots R_{\alpha_{2n-1}\,\alpha_{2n}}
{}^{\beta_{2n-1}\,\beta_{2n}},
\end{equation}
where also $R^{(1)}_{\alpha\beta}\equiv R_{\alpha\beta}$ and
$R^{(1)}\equiv R$. With the usual definition of trace (i.e., the
standard contraction of any two indices), one obtains~\cite{farb}
${\sl trace}\, R^{(n)}_{\alpha\beta}/n=R^{(n)}$, hence one cannot
achieve a similar trace relation for the two parts of the whole
Lovelock tensor. Nevertheless, we have shown~\cite{farb} that with
the generalized trace (denoted by the Trace, as distinct from the
trace) operator, defined as follows, one can modify the original
form of the trace relation adequately, and achieves ${\sl Trace}\,
R^{(n)}_{\alpha\beta}=R^{(n)}$, hence ${\sl Trace}\, R^{\rm
(Lovelock)}_{\alpha\beta}=R^{\rm (Lovelock)}$.

For a general $\bigl({N\atop M}\bigr)$ tensor which is a
homogeneous function of degree $h$ with respect to the metric and
its derivatives (denoted in the brackets attached to the upper
left--hand side of the tensor), we have defined~\cite{farb}
\begin{equation}\label{Tracedu}
\textrm{Trace}\, {}^{[h]}A^{\alpha_1\ldots\alpha_N}
{}_{\beta_1\ldots\beta_M} :=\!\cases{
       \frac{1}{h-\frac{N}{2}+\frac{M}{2}}\, \textrm{trace}\,
       {}^{[h]}A^{\alpha_1\ldots\alpha_N}{}_{\beta_1\ldots\beta_M} &
       \textrm{when $h-\frac{N}{2}+\frac{M}{2}\not=0$}\cr
       \textrm{trace}\, {}^{[h]}A^{\alpha_1\ldots\alpha_N}
       {}_{\beta_1\ldots\beta_M} &
       \textrm{when $h-\frac{N}{2}+\frac{M}{2}=0$,}\cr}
\end{equation}
where, without loss of generality, the homogeneity degree number
({\bf HDN}) conventions of each of $g^{\mu\nu}$ and
$g^{\mu\nu}{}_{,\alpha}$ are taken to be one. The HDN of a term
consisted of cross functions is obviously found by adding the HDN
of each of the cross functions. Hence, for example for two
specified cross functions with the HDN $h'$ and $h$, when
$h'+h\neq -1$ and $h'\not=0$, one gets
\begin{equation}\label{Tracexampl}
\textrm{Trace}\,\bigl({}^{[h']}C\,
{}^{[h]}A_{\mu\nu}\bigr)=\cases{
     \frac{h+1}{h'+h+1}\, {}^{[h']}C\,\textrm{Trace}\,
     {}^{[h]}A_{\mu\nu} &
     \textrm{for $h\not=-1$}\cr
     \frac{1}{h'}\, {}^{[h']}C\,\textrm{Trace}\,
     {}^{[h]}A_{\mu\nu} &
     \textrm{for $h=-1$,}\cr}
\end{equation}
and when $h'+h= -1$ and $h'\not=0$, one has
\begin{equation}\label{Tracexamp2}
{\rm Trace}\,\Bigl({}^{[h']}C\, {}^{[h]}A_{\mu\nu}\Bigr)
=\bigl(h+1\bigr)\, {}^{[h']}C\,{\rm Trace}\, {}^{[h]}A_{\mu\nu}\,
.
\end{equation}

Note that, as a homogeneous Euler--Lagrange expression has a
uniform HDN, then one can work with the usual trace instead of the
generalized trace operator. However, the notion of the generalized
trace operator has been introduced to be effective when one
considers the Einstein--Hilbert Lagrangian plus higher order terms
as a complete gravitational Lagrangian, i.e. when one works with
an inhomogeneous Lagrangian constructed linearly in terms of
homogeneous terms.

\section{Third Order Lagrangian Terms}
\indent

The third order Lagrangian of the Lovelock Lagrangian
is~\cite{mulhc,whelt,fard}
\begin{equation}\label{loveiii}
L^{(3)}={1 \over\kappa^2}\left(K_1-12 K_2+3 K_3+16 K_4+24 K_5-24
K_6+2 K_7-8 K_8\right),
\end{equation}
where\footnote{Any other relevant term can easily be written in
terms of these eight terms, e.g.
\begin{equation}\label{egk8}
R^{\sigma\tau}{}_{\mu\nu}R^{\mu\lambda}{}_{\sigma\rho}
R^\nu{}_{\lambda\tau}{}^\rho=K_7/4+K_8.
\end{equation}
Actually, according to Ref.~\cite{fkwc92}, the dimension of the
basis of local cubic invariants with the Riemann tensor (without
derivatives) is eight for $D>5$ dimensions.}
 {\setlength\arraycolsep{24pt}
\begin{eqnarray}\label{kscon}
 &K_1\equiv R^3,\qquad\qquad\qquad\ \
 &K_2\equiv RR_{\mu\nu}R^{\mu\nu},\qquad\qquad\quad\ \
  K_3\equiv RR_{\rho\tau\mu\nu}R^{\rho\tau\mu\nu},\ \cr
 &K_4\equiv R^{\mu\nu}R_{\mu\gamma}R_{\nu}{}^{\gamma}\,,\qquad\ \
 &K_5\equiv R_{\rho\tau}R_{\mu\nu}R^{\rho\mu\tau\nu},\qquad\qquad
  K_6\equiv R_{\lambda\rho}R^{\lambda\tau\mu\nu}R^{\rho}{}_{\tau\mu\nu}\,,\ \cr
 &K_7\equiv R^{\sigma\tau}{}_{\mu\nu}R^{\mu\nu}{}_{\lambda\rho}
  R^{\lambda\rho}{}_{\sigma\tau}\,,
 &K_8\equiv R^{\sigma\tau}{}_{\mu\nu}R^{\mu\lambda}{}_{\sigma\rho}
  R^{\nu\rho}{}_{\tau\lambda}.\quad
\end{eqnarray}  }
These third order terms are the only linearly independent scalar
terms, and the corresponding third order generic Lagrangian can be
written as
\begin{equation}\label{liii}
L^{(3)}_{\rm generic}={1 \over\kappa^2}\left(b_1 K_1+b_2 K_2+b_3
K_3+b_4 K_4+b_5 K_5+b_6 K_6+b_7 K_7+b_8 K_8\right),
\end{equation}
where the $b_i$'s are arbitrary dimensionless constants, and
obviously, in six dimensions, only seven of these eight terms are
effective (see identity (\ref{eeight})). From definition
(\ref{lovet}), the third term of the Lovelock tensor
is~\cite{mulhc,fard}
\begin{eqnarray}\label{giii}
&G^{(3)}_{\alpha\beta}=3\biggl\{\!\!\!\!
             &R^2\, R_{\alpha\beta}-4\Bigl(
              R\, R_{\alpha\mu}\, R_{\beta}{}
              ^{\mu}+R\, R_{\alpha\mu\beta\nu}\,
              R^{\mu\nu}+R_{\alpha\beta}\,
              R_{\mu\nu}R^{\mu\nu}\Bigr)\cr
           &&+\Bigl(2R\, R_{\alpha\rho\mu\nu}\, R_{\beta}{}^{\rho\mu\nu}+R_{\alpha\beta}\, R_{\rho\tau\mu\nu}\,
             R^{\rho\tau\mu\nu}\Bigr)+8\Bigl(R_{\alpha\mu}\,
             R_{\beta\nu}\, R^{\mu\nu}\cr
           &&-\!R_{\alpha\mu\nu\beta}\,
             R^{\mu\rho} R^{\nu}{}_{\rho}\Bigr)\!-8\left[\Bigl(R_{\alpha\mu}\,
             R_{\rho\tau}\, R_{\beta}{}
             ^{\rho\tau\mu}+\ \alpha\leftrightarrow\beta\Bigr)
             -\!R_{\rho\alpha\beta\tau}\,
             R^{\rho\mu\nu\tau}R_{\mu\nu}\right]\cr
          &&-4\Bigl[\Bigl(R_{\alpha\mu}\, R_{\beta
            \nu\rho\tau}\, R^{\mu\nu\rho\tau}
            +\ \alpha\leftrightarrow\beta\Bigr)
            +R_{\alpha\mu\rho\tau}\, R_{\beta
            \nu}{}^{\rho\tau}\, R^{\mu\nu}\cr
          &&+2R_{\alpha\rho\tau\mu}\, R_{\beta}{}^{\rho\tau}{}_{\nu}\,
            R^{\mu\nu}+R_{\alpha\mu\beta\nu}\,
            R^{\mu\rho\tau\lambda}\,
            R^{\nu}{}_{\rho\tau\lambda}\Bigr]
            +2R_{\alpha\lambda\mu\nu}\, R_{\beta}{}^{\lambda}{}
            _{\rho\tau}\, R^{\mu\nu\rho\tau}\cr
          &&+8R_{\alpha\mu\nu\lambda}\,
            R_{\beta\rho\tau}{}^{\lambda}\,
            R^{\mu\tau\rho\nu}\biggr\}-{1\over 2}\, g_{\alpha\beta}\> \kappa
            ^{2}\, L^{(3)}.
\end{eqnarray}
After some rather bulky calculations, the full appearance for the
Euler--Lagrange expression of the third order generic Lagrangian
can be written either as
\begin{equation}\label{nliii}
G^{(3a)}_{({\rm generic})\alpha\beta} =3\Bigl(N_{\alpha\beta}
+H_{\alpha\beta}\Bigr)-{1\over 2} g_{\alpha\beta}\,
\Bigl(\kappa^{2}\,L^{(3)}_{\rm generic}+M^{(3)}\Bigr),
\end{equation}
or, inspired from the Euler--Lagrange expression (\ref{defggrav})
and assumptions
\begin{equation}\label{asume}
\delta L^{(\rm gravi.)}/\delta g^{\alpha\beta}\equiv R^{(\rm
gravi.)}_{\alpha\beta}/\kappa^2\qquad {\rm and}\qquad L^{(\rm
gravi.)}\equiv R^{(\rm gravi.)}/\kappa^2,
\end{equation}
as
\begin{equation}\label{nliiia}
G^{(3b)}_{({\rm generic})\alpha\beta} =3\Bigl(N_{\alpha\beta}
+H_{\alpha\beta} -{1\over 6}g_{\alpha\beta}\, M^{(3)}
\Bigr)-{1\over 2}\kappa^2\, g_{\alpha\beta}\, L^{(3)}_{\rm
generic}\, ,
\end{equation}
where\footnote{Also, see relation (\ref{a1}).}\
 \cite{fard}
\begin{eqnarray}\label{mlt}
&M^{(3)}\!\equiv\!\!\!\!
    &\Bigl(\!-3b_4+2b_5\Bigr)\!\Bigl(K_4-K_5\Bigr)
     +\Bigl(b_5+b_6\Bigr)\Bigl(K_6-K_7-2 K_8\Bigr)\cr
   &&-\Bigl(12b_1+b_2\Bigr)K_9-\Bigl(4b_2+2b_5\Bigr)K_{10}-\Bigl(4b_3+{1\over 2}b_5+b_6\Bigr)
     K_{11}\cr
   &&-\Bigl(2b_2+3b_4-b_5\Bigr)K_{12}-\Bigl(3b_4-4b_5-2b_6\Bigr)K_{13}
     -\Bigl(12b_1+2b_2+{3\over 4}b_4\Bigr)K_{15}\cr
   &&-\Bigl(4b_2+4b_5+2b_6\Bigr)K_{16}-\Bigl(4b_3+{1\over 2}b_6\Bigr)K_{17}\, ,
\end{eqnarray}
\begin{eqnarray}\label{nlt}
&N_{\alpha\beta}\equiv\!\!\!\!
  &b_1 R^2R_{\alpha\beta}-{4\over 3}b_3
   RR_{\alpha\rho}R_{\beta}{}^{\rho}
   +{2\over 3}\Bigl(b_2+2b_3\Bigr)RR^{\mu\nu}R_{\alpha\mu\beta\nu}+{1\over 3}b_2
   R_{\alpha\beta}R^{\mu\nu}R_{\mu\nu}\cr
 &&+{2\over 3}b_3 RR_{\alpha\lambda\rho\sigma}R_{\beta}{}^{\lambda\rho\sigma}
   +{1\over 3}b_3 R_{\alpha\beta}
   R^{\lambda\rho\sigma\tau}R_{\lambda\rho\sigma\tau}-{1\over 3}
   \Bigl(b_5+2b_6\Bigr)R_{\alpha\mu}R_{\beta\nu}
   R^{\mu\nu}\cr
 &&-\!\Bigl({2\over 3}b_5+{2\over 3}b_6-b_8\Bigr)
   R_{\alpha\mu\nu\beta}
   R^{\mu\rho}R^{\nu}{}_{\rho}-\!\Bigl({1\over 2}b_4+{1\over 3}b_5+{1\over 3}
   b_6\Bigr)\!\biggl(R_{\alpha\mu}R_{\lambda\rho}R_{\beta}{}^{\lambda\rho\mu}\cr
 &&+\ \alpha\leftrightarrow\beta \biggr)
   +\Bigl(-{2\over 3}b_6+b_8\Bigr)R_{\lambda\alpha\beta\rho}R^{\lambda\mu\nu\rho}R_{\mu\nu}-2b_7\biggl(%
   R_{\alpha\mu}R_{\beta\lambda\rho\sigma}
   R^{\mu\lambda\rho\sigma}\cr
 &&+\ \alpha\leftrightarrow\beta \biggr)
   +\Bigl({2\over 3}b_6+6b_7\Bigr)R_{\alpha\mu\lambda\rho}
   R_{\beta\nu}{}^{\lambda\rho}R^{\mu\nu}
   +\Bigl({2\over 3}b_6-b_8\Bigr)R_{\alpha\lambda\rho\mu}R_{\beta}{}^{\lambda\rho}{}_{\nu}R^{\mu\nu}\cr
 &&+\!\Bigl({1\over 3}b_6-{1\over 2}b_8\Bigr)R_{\alpha\mu\beta\nu}R^{\mu\lambda\rho\sigma}
   R^{\nu}{}_{\lambda\rho\sigma}+\!\Bigl(-{2\over
   3}b_6-b_7+{3\over 2}b_8\Bigr)R_{\alpha\tau\mu\nu}R_{\beta}{}^{\tau}
   {}_{\lambda\rho}R^{\mu\nu\lambda\rho}\cr
 &&+\Bigl({4\over 3}b_6+4b_7-4b_8\Bigr)R_{\alpha\mu\nu\tau}
   R_{\beta\lambda\rho}{}^{\tau}R^{\mu\rho\lambda\nu}
\end{eqnarray}
and
\begin{eqnarray}\label{hlt}
&H_{\alpha\beta}\!\equiv \!\biggl[\!\!\!\!\!
  &-\Bigl(b_1+{1\over 6}b_2+{1\over 24}b_5\Bigr)R_{;\,\alpha}
   R_{;\,\beta}
   +{1\over 3}\Bigl(b_2+4b_3+{1\over 2}b_5+{1\over 2}
   b_6\Bigr)R^{;\,\rho}R_{\alpha\beta ;\,\rho}\cr
 &&\!-\!\Bigl(b_1\!+\!{1\over 6}b_2\!+\!{1\over 3}b_3\Bigr)%
   RR_{;\,\alpha\beta}\!
   +{1\over 6}\Bigl(b_2\!+\!4b_3\Bigr)
   RR_{\alpha\beta ;\,\rho}{}^\rho
   \!+{1\over 6}\Bigl(b_2\!+\!{1\over 2}b_5\Bigr)\biggl(
   R_{;\,\rho}{}^\rho R_{\alpha\beta}\cr
 &&-2R^{\mu\nu}R_{\mu\nu;\,\alpha\beta}\biggr)
   -\Bigl({2\over 3}b_3+{1\over 6}b_6-{1\over 4}b_8\Bigr)\biggl(
   R^{;\,\mu\nu}R_{\alpha\mu\nu\beta}
   +{1\over 2}
   R_{\lambda\rho\sigma\tau;\,\alpha\beta}
   R^{\lambda\rho\sigma\tau}\biggr)\cr
 &&\!-\Bigl({1\over 3}b_2+\!{1\over 3}b_5+\!{1\over 2}b_8\Bigr)
   R_{\mu\nu;\,\alpha}R^{\mu\nu}{}_{;\,\beta}
   +\!\Bigl({1\over 2}b_4+\!{2\over 3}b_6+\!2b_7-{1\over 2}b_8\Bigr)%
   R_{\alpha\mu;\,\nu}R_{\beta}{}^{\mu;\,\nu}\cr
 &&-\Bigl({1\over 6}b_5+{1\over 3}b_6+2b_7\Bigr)R_{\alpha\mu;\,\nu}
   R_{\beta}{}^{\nu;\,\mu}
   -\Bigl({1\over 3}b_3+{1\over 8}b_8\Bigr)R_{\lambda\rho\sigma\tau;\,\alpha}
   R^{\lambda\rho\sigma\tau}{}_{;\,\beta}\cr
 &&\!+\!{1\over 3}\Bigl(b_5+b_6\Bigr)\!\biggl(
   R^{\mu\nu}R_{\alpha\beta ;\,\mu\nu}
   -R^{\mu\nu;\,\rho}R_{\alpha\mu\rho
   \beta ;\,\nu}\biggr)
   -\Bigl({1\over 3}b_5+b_8\Bigr)\!\biggl(
   {1\over 2}R_{\mu\nu;\rho}{}^{\rho}R_{\alpha}{}^{\mu\nu}{}_{\beta}\cr
 &&-\!R_{\mu\nu;\,\alpha\rho}R_{\beta}
   {}^{\mu\nu\rho}\biggr)
   +\!\Bigl({1\over 6}b_6+\!b_7-\!{1\over 4}b_8\Bigr)\!\biggl(
   R_{\lambda\rho\sigma\alpha;\,\tau}R^{\lambda\rho\tau}{}_{\beta}{}^{;\,\sigma}
   -4R_{\alpha\rho;\,\mu\nu}R_{\beta}{}
   ^{\mu\nu\rho}\biggr)\cr
 &&+\Bigl({1\over 3}b_6-b_8\Bigr)R^{\mu\nu;\,\rho}R_{\alpha\mu\nu\rho;\,
   \beta}
   -\Bigl({1\over 3}b_2+{4\over 3}b_3+{1\over 4}b_4+{1\over 6}b_6\Bigr)
   R^{;\,\rho}R_{\alpha\rho;\,\beta}\cr
 &&-\Bigl({1\over 3}b_2+{1\over 4}b_4
   +{1\over 6}b_5+{1\over 6}b_6\Bigr)R_{;\,\alpha}{}^{\rho}
   R_{\beta\rho}
   -\Bigl({1\over 2}b_4-{1\over 3}b_5+{1\over 3}b_6-b_8\Bigr)\times\cr
 &&R_{\alpha\mu;\,\nu}R^{\mu\nu}{}_{;\,\beta}
   +\Bigl({1\over 2}b_4+{1\over 3}b_6\Bigr)\biggl(
   R_{\alpha}{}^{\rho}
   R_{\beta\rho;\,\mu}{}^{\mu}
   -R^{\mu\nu}R_{\alpha\mu;\,\beta\nu}\biggr)
   \biggr]
   +\ \alpha\leftrightarrow\beta .
\end{eqnarray}
As seen, in contrast to the $G^{(3)}_{\alpha\beta}$, the
$G^{(3)}_{({\rm generic})\alpha\beta}$ is up to the fourth order
jet--prolongation of the metric, as expected\rlap.\footnote{For a
few solutions (mainly the black hole solutions) to the
curvature--cubed (sometimes also called the six derivative)
interactions, see, e.g., Refs.~\cite{MaTe06OlRa10MyRo10}.}\
 The third and fourth order terms are due
to the $H_{\alpha\beta}$ and $M^{(3)}$ terms, which will vanish if
and only if the constant coefficients satisfy the ratios
\begin{eqnarray}\label{cliii}
 &b_2=-12\, b_1,\qquad
 &b_3=3\, b_1,\qquad b_4=16\, b_1,\qquad
  b_5=24\, b_1, \cr
 &b_6=-24\, b_1,\qquad
 &b_7=2\, b_1,\qquad   b_8=-8\, b_1.
\end{eqnarray}
These conditions are exactly the ratios of the constituent
coefficients of the special case of $L^{(3)}$ that leads to the
$G^{(3)}_{\alpha\beta}$. Also, with the above ratios, the
$N_{\alpha\beta}$ term will be equal to the corresponding
counterpart of the $G^{(3)}_{\alpha\beta}$.

To amend the Lagrangian of sixth order gravity~\cite{goss}, Berkin
and {\it et al.}~\cite{berm} discussed that the Lagrangian term of
$R\,\Square\, R$ is a third order Lagrangian based on the
dimensionality scale;\footnote{Since two derivatives are
dimensionally equivalent to one Riemann--Christoffel tensor or any
one of its contractions.}\
 however, it can be better justified on account of its HDN which is three.
Indeed, to classify different gravitational Lagrangian terms, it
is straightforward to relate~\cite{farb,fare} the order $n$ in any
Lagrangian (as in the $L^{(n)}$) to represent its HDN and
referring to Lagrangians with their HDNs rather than their orders.
Hence, by gathering terms with the same HDN under one Lagrangian
label, in addition to the eight linearly independent terms, in
$D>5$ dimensions, which appear in the third order of the Lovelock
Lagrangian (and are up to the second order jet--prolongation of
the metric), there are~\cite{fard}, in general, another nine
linearly independent scalar terms, constructed from the
Riemann--Christoffel tensor and its contractions, with the HDN
three which are up to the third or even higher order
jet--prolongation of the metric. They all also satisfy the
dimensionality scale, and are\footnote{See also
Refs.~\cite{bpb,fkwc92,DecFol07}.}
 {\setlength\arraycolsep{27pt}
\begin{eqnarray}\label{ksconb}
K_9\equiv R\,\Square R\qquad\!
  &K_{10}\equiv R_{\mu\nu}\,\Square R^{\mu\nu}\quad\!\!\!
  &K_{11}\equiv R_{\mu\nu\rho\tau}\,\Square R^{\mu\nu\rho\tau}\cr
K_{12}\equiv R_{\mu\nu}R^{;\,\mu\nu}\,
  &K_{13}\equiv R_{\mu\nu ;\,\rho}R^{\mu\rho ;\,\nu}
  &K_{14}\equiv \Square\,^2 R\cr
K_{15}\equiv R_{;\,\rho}R^{;\,\rho}\quad
  &K_{16}\equiv R_{\mu\nu ;\,\rho}R^{\mu\nu ;\,\rho}
  &K_{17}\equiv R_{\mu\nu\rho\tau ;\,\alpha}R^{\mu\nu\rho\tau ;\,\alpha}
   \, ,
\end{eqnarray}  }
where $\Square\equiv {}_{;\,\rho}{}^{\rho}$. Each of the $K_i$'s
gives a dimensionless action in six dimensions. Any other relevant
term, e.g. $R_{\mu\nu ;\,\rho\tau}R^{\rho\mu\nu\tau}$, can be
written in terms of these terms, see, e.g. the last relation of
(\ref{useful11}). Besides, not all of their corresponding
Euler--Lagrange expressions are independent; however, see the
Appendix~A for the effects of these nine terms as scalar
Lagrangians.

\section{$L^{(3)}_{\rm generic}$ With The Trace Property}
\indent

In this section, we investigate the analogy for the field tensor
of the $L^{(3)}_{\rm generic}$, and then, in the next section, for
any type of third order Lagrangian including those mentioned
in~(\ref{ksconb}). The relevant Euler--Lagrange expression
$G^{(3)}_{({\rm generic})\alpha\beta}$ has been written as
relations (\ref{nliii}) and (\ref{nliiia}); however, we also
demand to have them as in~(\ref{drFg}), i.e. $G^{(3)}_{({\rm
generic})\alpha\beta}=R^{(3)}_{({\rm
generic})\alpha\beta}-g_{\alpha\beta}\, R^{(3)}_{\rm generic}/2$,
but propose to find out whether, and for what conditions, the
relation
\begin{equation}\label{rTiii}
{\rm Trace}\, R^{(3)}_{({\rm generic})\alpha\beta}= R^{(3)}_{\rm
generic}
\end{equation}
can be valid. We carry out this investigation for both appearances
of $G^{(3)}_{({\rm generic})\alpha\beta}$ in the following two
parts.
 \\ \\
{\bf Part (a):\ \ The Case $G^{(3a)}_{({\rm
                  generic})\alpha\beta}$}
\\

In this case, from (\ref{nliii}), we have $R^{(3a)}_{({\rm
generic})}=\kappa^{2}\,L^{(3)}_{\rm generic}+M^{(3)}$ and
\begin{equation}\label{trnh}
{\rm Trace}\, R^{(3a)}_{({\rm generic})\alpha\beta}=3\Bigl({\rm
Trace}\, N_{\alpha\beta}+{\rm Trace}\, H_{\alpha\beta}\Bigr).
\end{equation}
Using the definition of generalized trace and the fact that each
of the $N_{\alpha\beta}$ and $H_{\alpha\beta}$, and hence the
$R^{(3a)}_{({\rm generic})\alpha\beta}$, has the HDN
two~\cite{farb}, we get
\begin{eqnarray}
{\rm Trace}\, N_{\alpha\beta}={1\over
3}\biggl[\!\!\!\!\!\!\!\!\!\!
    &&b_1 K_1+b_2 K_2+b_3 K_3+\Bigl({1\over 3}b_5-b_8\Bigr)K_4
      +\Bigl(b_4+{2\over 3}b_5+b_8\Bigr)K_5\cr
  {}&&+\Bigl({5\over 3}b_6+2b_7-{3\over 2}b_8\Bigr)K_6
      +\Bigl(-{2\over 3}b_6-b_7+{3\over 2}b_8\Bigr)K_7
      +\Bigl(-{4\over 3}b_6-4b_7+4b_8\Bigr)K_8\biggr]
\end{eqnarray}
and
\begin{eqnarray}
{\rm Trace}\, H_{\alpha\beta}={1\over
3}\biggl[\!\!\!\!\!\!\!\!\!\!
         &&\Bigl({1\over 3}b_5-{2\over 3}b_6-4b_7+2b_8\Bigr)
           \Bigl(K_6-K_7-2\, K_8\Bigr)
           -\Bigl(2b_1-{1\over 3}b_2-{2\over 3}b_3\cr
       {}&&-{1\over 6}b_5\Bigr)K_9
           +\Bigl(-{2\over 3}b_2+b_4+{2\over 3}b_6+b_8\Bigr)K_{10}
           +\Bigl(-{2\over 3}b_3-{1\over 6}b_5+{1\over 6}b_6\cr
       {}&&+2b_7-{3\over 4}b_8\Bigr)K_{11}
           -\Bigl({2\over 3}b_2-{4\over 3}b_3+b_4-{1\over 3}b_5
           -{1\over 3}b_6+{1\over 2}b_8\Bigr)K_{12}\cr
       {}&&-\Bigl(b_4-b_5+4b_7\Bigr)K_{13}
           +\Bigl(-2b_1+{4\over 3}b_3-{1\over 4}b_4+{1\over 4}
           b_5+{1\over 6}b_6\Bigr)K_{15}\cr
       {}&&-\!\Bigl({2\over 3}b_2-b_4+{2\over 3}b_5-{2\over 3}b_6
           -4b_7\!\Bigr)K_{16}\! +\!\Bigl(\!-{2\over 3}b_3+{1\over 6}b_6+b_7
           -{1\over 2}b_8\!\Bigr)K_{17}\!\biggr].
\end{eqnarray}
Then, after all the necessary substitutions have been made, we
find that the equality of (\ref{rTiii}) holds if and only if the
following conditions between those eight non--zero constituent
coefficients of $L^{(3)}_{\rm generic}$ are satisfied. That is,
the constant coefficients are~not all independent but obey the
constraint
 {\setlength\arraycolsep{27pt}
\begin{eqnarray}\label{cnliii}
  &b_4=2(-b_2+4b_3)/3,\qquad\qquad\ \
  &b_5=-60\, b_1-8\, b_2-4\, b_3,\cr
  &b_6=30\, b_1+2\, b_2-10\, b_3,\quad\qquad\,\,
  &b_7=30\, b_1+14b_2/3+28b_3/3,\cr
  &b_8=100\, b_1+12\, b_2+12\, b_3.\,\qquad &
\end{eqnarray}  }
\indent
 These conditions are satisfied for (as must be the case)
the non--generic Lagrangian $L^{(3)}$, i.e. for the coefficients
given by relation (\ref{cliii})\rlap.\footnote{Just by choosing
$b_2=-12\, b_1$ and $b_3=3\, b_1$ (as in (\ref{cliii})),
constraint (\ref{cnliii}) then gives the remaining coefficients
with the same ratios as those of (\ref{cliii}).}\
 Though, it is~not only this Lagrangian that has the analogy of the Einstein
tensor and, using constraint (\ref{cnliii}), there are more
combinations for $L^{(3)}_{\rm generic}$ with {\it three} degrees
of freedom (out of its eight coefficients) that have the required
trace property (see below for other examples).

As mentioned, the $L^{(3)}_{\rm generic}$ is a Lagrangian which
gives $G^{(3a)}_{({\rm generic})\alpha\beta}$ as in relation
(\ref{nliii}) with $R^{(3a)}_{\rm generic}=\kappa^2\,L^{(3)}_{\rm
generic}+M^{(3)}$; however, one can define\footnote{Note that,
although this $L^{\prime\, (3)}_{\rm generic}$ is again up to the
fourth order (and not up to the second order) jet--prolongation of
the metric, but its HDN is still three, the same as the
$L^{(3)}_{\rm generic}$ term, in agreement with our demand of
gathering terms with the same HDN under one Lagrangian label.}
\begin{equation}\label{liiipri}
\kappa^2\,L^{\prime\, (3)}_{\rm generic}\equiv
\kappa^2\,L^{(3)}_{\rm generic}+M^{(3)},
\end{equation}
where, as is shown in the Appendix~A, the term $M^{(3)}$ is a
complete divergence (see relation~(\ref{a1})) and will give no
contribution to the variation of the relevant action. Therefore,
this new Lagrangian, $L^{\prime\, (3)}_{\rm generic}$, gives the
same $G^{(3a)}_{({\rm generic})\alpha\beta}$ with $R^{(3a)}_{\rm
generic}\equiv \kappa^2\,L^{\prime\, (3)}_{\rm generic}$ which is
analogous to the appearances of the case $L^{(3)}$, the complete
Lovelock Lagrangian (relation (\ref{rrlove})) and the assumption
(\ref{asume}).

If one wants to use different choices of $L^{(3)}_{\rm generic}$,
for which more than {\it three} of the constituent coefficients
are zero (as, for example, not all of the terms in $L^{(3)}_{\rm
generic}$ have been given by the superstring theory\footnote{See,
for example, Ref.~\cite{metsketo}.}),
 then the trace condition
(\ref{cnliii}) will~not be satisfied. Because, the maximum number
of coefficients that it permits at a time to be equal to zero
(otherwise all of them will be zero) are generally two, and only
for the following special cases are three. Therefore, those
choices of $L^{(3)}_{\rm generic}$ that miss more than the
permitted terms cannot provide the trace relation. The following
two cases are the only cases in which three coefficients of the
$L^{(3)}_{\rm generic}$ can simultaneously be made zero, though
there is still one degree of freedom remaining to satisfy the
trace condition. If one chooses $b_2=0=b_3$, then
constraint~(\ref{cnliii}) will give
\begin{equation}
  b_4=0,\ \qquad
  b_5=-60\, b_1,\ \qquad
  b_6=30\, b_1,\ \qquad
  b_7=30\, b_1,\ \qquad
  b_8\!=\! 100 b_1,
\end{equation}
and if one sets $b_2=-20 b_1/3$\ and $b_3=-5 b_1/3$, it then will
read
\begin{equation}
  b_4=0,\ \qquad
  b_5=0,\ \qquad
  b_6=100 b_1/3,\ \qquad
  b_7=-50 b_1/3,\ \qquad
  b_8=0.
\end{equation}

Similar to the Lagrangian constructed from the square of the Weyl
tensor, relation~(\ref{wsd}), there are only
two\footnote{According to Ref.~\cite{Erdmenger97} and the appendix
of Ref.~\cite{bgzv}, the dimension of the basis of local cubic
Weyl--invariants is two for $D>5$ dimensions, as, e.g., in {\it
four} dimensions, one has $A_1=4\, A_2$ since $5\,
C^{\sigma\tau}{}_{[\mu\nu}C^{\mu\nu}{}_{\lambda\rho}
C^{\lambda\rho}{}_{\sigma]\tau}=A_1-4\, A_2$.}\
 linearly independent scalars formed by the cubic of the Weyl
tensor. We choose the
$C^{\sigma\tau}{}_{\mu\nu}C^{\mu\nu}{}_{\lambda\rho}
C^{\lambda\rho}{}_{\sigma\tau}$ and
$C^{\sigma\tau}{}_{\mu\nu}C^{\mu\lambda}{}_{\sigma\rho}
C^{\nu\rho}{}_{\tau\lambda}$ (with the indices similar to the ones
of the $K_7$ and $K_8$ terms, respectively). Their expressions in
$D\geq 3$ dimensions (also, see the end of the Useful Relations in
the Appendix~A) are
\begin{eqnarray}\label{cwtd}
A_1\equiv C^{\sigma\tau}{}_{\mu\nu}C^{\mu\nu}{}_{\lambda\rho}
C^{\lambda\rho}{}_{\sigma\tau}=\!\!\!\!\!\!\!\!
 &&{8(2D-3)\over (D-2)^3(D-1)^2}\, K_1-{24(2D-3)\over (D-2)^3(D-1)}\,
   K_2+{6\over (D-2)(D-1)}\, K_3\cr
 &&+{16(D-1)\over (D-2)^3}\, K_4+{24\over (D-2)^2}\,
   K_5-{12\over (D-2)}\, K_6+K_7
\end{eqnarray}
and
\begin{eqnarray}\label{cwtda}
A_2\equiv C^{\sigma\tau}{}_{\mu\nu}C^{\mu\lambda}{}_{\sigma\rho}
C^{\nu\rho}{}_{\tau\lambda}=\!\!\!\!\!\!\!\!
 &&-{D^2+5D-10\over (D-2)^3(D-1)^2}\, K_1
   +{3(D^2+5D-10)\over (D-2)^3(D-1)}\, K_2-{3\over (D-2)(D-1)}\, K_3\cr
 &&-{2(5D-6)\over (D-2)^3}\, K_4-{3(D+2)\over (D-2)^2}\,
   K_5+{6\over (D-2)}\, K_6+K_8.
\end{eqnarray}
The coefficients of the above expressions can satisfy
constraint~(\ref{cnliii}) only if $D=6$. Their values in this
dimension respectively are\footnote{These values already are
consistent with when the generalization of the Gauss--Bonnet
theorem~\cite{cheachebkonospiv} in six dimensions has also been
considered.}
 {\setlength\arraycolsep{6pt}
\begin{eqnarray}\label{cwts}
b_1=9/200,\qquad
   &b_2=-27/40,\qquad
   &b_3=3/10,\qquad
    b_4=5/4,\cr
b_5=3/2,\qquad\quad
   &b_6=-3,\qquad\qquad
   &b_7=1,\qquad\quad\ \,
    b_8=0
\end{eqnarray}  }
and
 {\setlength\arraycolsep{6pt}
\begin{eqnarray}\label{cwtsa}
b_1=-7/200,\qquad
   &b_2=21/40,\qquad
   &b_3=-3/20,\qquad
    b_4=-3/4,\cr
b_5=-3/2,\qquad\quad
   &b_6=3/2,\qquad\quad
   &b_7=0,\qquad\qquad\,
    b_8=1.
\end{eqnarray}  }
Alternatively, one can obviously write expression~(\ref{cwtda}),
in six dimensions, effectively (using identity~(\ref{eeight}))
with the values
 {\setlength\arraycolsep{6pt}
\begin{eqnarray}\label{cwtsa2}
b_1=9/100,\qquad
   &b_2=-39/40,\qquad
   &b_3=9/40,\qquad
    b_4=5/4,\cr
b_5=3/2,\qquad\quad
   &b_6=-3/2,\qquad\quad
   &b_7=1/4,\qquad\,\
    b_8=0.
\end{eqnarray}  }
\indent
 The Lagrangian densities made by these two {\it cubic}
constructions of the Weyl tensor are the only conformal (Weyl)
invariant in {\it six} dimensions. However, according to
Refs.~\cite{desch,FefGra85}, in general, there are three linearly
independent Weyl--invariant combinations in six dimensions, and
two of them are obviously the above purely algebraic ones. The
another one (given by\footnote{However, see also
relations~(\ref{ebonora}), (\ref{a4kmm}) and~(\ref{earakelb}).}\
 relation~(\ref{edeser})) contains terms from the other nine third order
Lagrangian ones mentioned in~(\ref{ksconb}), and it is one of the
reasons why we will extend the analogy further to consider any
type of third order Lagrangian, including the terms
in~(\ref{ksconb}), in the next section.

In six dimensions, by the generalization of the Gauss--Bonnet
theorem, the $L^{(3)}\sqrt{-g}$ corresponds to the Euler
densities\footnote{See, for example, Ref.~\cite{bpb}. }\
 (i.e., it is a complete divergence), and hence one gets one more constraint
among the coefficients $b_1$ to $b_8$ in this dimension. That is,
in six dimensions, one has
\begin{equation}\label{eeight}
\left(G^{(K_1)}_{\alpha\beta}-12\, G^{(K_2)}_{\alpha\beta}+3\,
G^{(K_3)}_{\alpha\beta}+16\, G^{(K_4)}_{\alpha\beta}+24\,
G^{(K_5)}_{\alpha\beta}-24\, G^{(K_6)}_{\alpha\beta}+2\,
G^{(K_7)}_{\alpha\beta}-8\, G^{(K_8)}_{\alpha\beta}\right){\Bigr
|}_{\rm 6-dim.}\!\!\equiv 0.
\end{equation}
Therefore, for the remaining seven linearly independent effective
terms, after substituting for one of the term according to
identity~(\ref{eeight}), constraint~(\ref{cnliii}) for the {\it
new} coefficients reduces to an effective one. For example, by
substituting for the $K_8$ term, the effective constraint is
 {\setlength\arraycolsep{27pt}
\begin{eqnarray}\label{cnliiis}
  &b_1=-3(b_2+b_3)/25,\qquad\
  &b_4=2(-b_2+4b_3)/3,\cr
  &b_5=4(-b_2+4b_3)/5,\qquad\ \,
  &b_6=-4(2b_2+17b_3)/5,\cr
  &b_7=2(8b_2+43b_3)/15,\qquad\!\!
 \end{eqnarray}  }
with two degrees of freedom, where here we have chosen the $b_2$
and $b_3$ coefficients. Constraint~(\ref{cnliiis}) is satisfied by
the values given in~(\ref{cwts}) and~(\ref{cwtsa2}), as expected.
 \\ \\
{\bf Part (b):\ \ The Case $G^{(3b)}_{({\rm
generic})\alpha\beta}$}
\\

Now, we calculate the generalized trace of the first part of
relation (\ref{nliiia}), knowing the fact that the HDN of the
$M^{(3)}$ is three. By using relation (\ref{Tracexampl}), we get
\begin{equation}\label{trnha}
{\rm Trace}\, R^{(3b)}_{({\rm generic})\alpha\beta}=3\Bigl({\rm
Trace}\, N_{\alpha\beta}+{\rm Trace}\, H_{\alpha\beta}\Bigr)
-{D\over 6}\, M^{(3)},
\end{equation}
and again, after all the necessary substitutions and calculations
have been performed, we find that the equality of (\ref{rTiii}),
for non--zero coefficients, holds if and only if
\begin{eqnarray}\label{cnliiia}
 &&b_4=2(-b_2+4b_3)/3,\cr
 &&b_5=-12(D-1)b_1-(D+2)b_2-4\, b_3,\cr
 &&b_6=6(D-1)b_1+2\, b_2-2(D-1)b_3,\cr
 &&b_7=3(D-2)(D-1)b_1/2+(D-2)(D+1)b_2/6
   +(D^2+5\, D-10)b_3/6,\cr
 &&b_8=4(D-1)^2b_1+D^2\, b_2/3+4(2\,D-3)b_3/3,
\end{eqnarray}
where at least\footnote{In three dimensions, there are other
identities, e.g. relations (\ref{weylcub3d1}) and
(\ref{weylcub3d2}), that should also be considered; see the
appendix of Ref.~\cite{bgzv} too.}\
 $D> 3$. Though constraint (\ref{cnliiia})
depends on the dimension of space--time, it still gives three
degrees of freedom out of the eight coefficients. It is satisfied
by the $L^{(3)}$ in {\it any} dimension. This can be obvious after
arranging constraint (\ref{cnliiia}) with respect to the
dimension, $D$, and its powers for each coefficient. Then from
these, one can easily find that the {\it only}\ combination which
is independent of the dimension is that of the $L^{(3)}$.

Also, the coefficients of the scalar Lagrangians made by the cubic
of the Weyl tensor, relations~(\ref{cwtd}) and~(\ref{cwtda}),
identically satisfy constraint~(\ref{cnliiia}) independent of the
dimension of space--time $D$.

As is evident, by comparing relations~(\ref{trnh})
and~(\ref{trnha}), constraint (\ref{cnliiia}) in six dimensions is
exactly the same as condition (\ref{cnliii}). The reason beyond it
is that, suppose, in general, a $P$ scalar term is the necessary
shifting term by which the two appearances, similar to part~(a)
and part~(b) (i.e., similar to relations (\ref{nliii}) and
(\ref{nliiia})), are arranged. Hence, by comparing these two
parts, e.g. for the relevant part similar to part~(b), one obtains
\begin{equation}
R^{(nb)}_{(\rm generic)\alpha\beta}\equiv R^{(na)}_{(\rm
generic)\alpha\beta}-g_{\alpha\beta}\,P\qquad\quad\ {\rm
and}\qquad\quad\ R^{(nb)}_{\rm generic}\equiv R^{(na)}_{\rm
generic}-2P.
\end{equation}
Now, in order to achieve ${\sl Trace}\,R^{(na)}_{(\rm
generic)\alpha\beta}=R^{(na)}_{\rm generic}$ and ${\sl
Trace}\,R^{(nb)}_{(\rm generic)\alpha\beta}=R^{(nb)}_{\rm
generic}$ simultaneously be satisfied, we must have ${\sl
Trace}\bigl(g_{\alpha\beta}\,P\bigr)=2P$. If $P$ is a homogeneous
function of degree $h$ with respect to the
metric\rlap,\footnote{Though, according to the idea of gathering
terms with the same HDN under one Lagrangian label, $P$ must be a
homogeneous function of degree $n$ for the $L^{(n)}_{\rm
generic}$.}\
 then, using the generalized trace definition (\ref{Tracedu}), one will get
\begin{equation}\label{trrprim}
\cases{D=2h&\quad when $h\not=0$\cr
        D=2&\quad when $h=0$.\cr}
\end{equation}
In the case of $L^{(2)}_{\rm generic}$, we have shown~\cite{fare}
that $P\propto \Square\, R$, hence it has $h=2$ (as expected), and
thus $D=4$. In the case of $L^{(3)}_{\rm generic}$, $P$ is
proportional to the $M^{(3)}$, therefore it has $h=3$ (again, as
expected), and thus $D=6$.

\section{Other Weyl Invariants}
\indent

In this section, we extend the analogy to include the other nine
third order Lagrangian terms in~(\ref{ksconb}). In addition to the
reason mentioned almost at the end of part~(a), this extension is
needed because, the conformal anomalies in six dimensions contain
the term $K_{14}$ (see relation~(\ref{egilkey})) which appears
only when the Lagrangian terms~(\ref{ksconb}) are considered, see,
e.g., relations~(\ref{a15}) and~(\ref{a17}). However, as explained
in the Appendix~A, it should suffice to consider the Lagrangian
terms $K_1$ to~$K_{10}$ (see relation~(\ref{a13})) rather than
the~$K_1$ to~$K_{17}$. Though, as will be illustrated below, in
order to be able to apply the trace analogy to an effective total
third order Lagrangian, one, in practice, needs to consider all of
the~$K_9$, $K_{10}$ and~$K_{11}$ terms simultaneously in such a
Lagrangian. Hence, by neglecting identity~(\ref{a21}), we consider
the most general effective expression for the total third order
Lagrangian to be\footnote{We have dropped the prime sign on the
$b_i$'s.}
\begin{equation}\label{lagtotal}
L^{(3)}_{\rm total}\mathrel{\mathop=^{\rm eff}}L^{(3)}_{\rm
generic}+(b_9K_9+b_{10}K_{10}+b_{11}K_{11})/\kappa^2.
\end{equation}
Furthermore, we again want to write its corresponding
Euler--Lagrange expression the same as~(\ref{drFg}), i.e.
$G^{(3a)}_{({\rm total})\alpha\beta} =R^{(3a)}_{({\rm
total})\alpha\beta}-g_{\alpha\beta}\, R^{(3a)}_{\rm total}/2$,
where\footnote{In this section, as we intend to employ the results
only for the six--dimensional case, we apply the analogy only for
the appearance of the $G^{(3a)}_{({\rm total})\alpha\beta}$ in
accord with part~(a) of the previous section. However, we also
perform the analogy for the appearance of the $G^{(3b)}_{({\rm
total})\alpha\beta}$, in accord with part~(b), in Appendix~B.}
\begin{equation}\label{dgtotal}
G^{(3a)}_{({\rm total})\alpha\beta}=G^{(3a)}_{({\rm
generic})\alpha\beta}+b_9\, G^{(K_9)}_{\alpha\beta} +b_{10}\,
G^{(K_{10})}_{\alpha\beta}+b_{11}\, G^{(K_{11})}_{\alpha\beta}
\end{equation}
and similar expressions for $R^{(3a)}_{({\rm total})\alpha\beta}$
and $R^{(3a)}_{\rm total}$. The $G^{(3a)}_{({\rm
generic})\alpha\beta}$ is given by relation~(\ref{nliii}) and the
Euler--Lagrange expressions of the~$K_9$ to $K_{11}$ Lagrangian
terms are given by relations~(\ref{a15})--(\ref{a18}). Also, we
propose to find out the conditions for which ${\sl Trace}\,
R^{(3a)}_{({\rm total})\alpha\beta}=R^{(3a)}_{\rm total}$.

After all the necessary substitutions and calculations, we find
that the trace relation, for non--zero coefficients, holds if and
only if
 {\setlength\arraycolsep{10pt}
\begin{eqnarray}\label{cgnliii}
 &b_5=-60\, b_1-122\, b_2/15-52\, b_3/15-b_4/5,\qquad
 &b_6=30\, b_1-4\, b_2/15-14\, b_3/15-17\, b_4/5,\cr
 &b_7=30\, b_1+32\, b_2/5+12\, b_3/5+13\, b_4/5,\qquad\quad
 &b_8=100\, b_1+40\, b_2/3+20\, b_3/3+2\, b_4,\cr
 &b_9=b_2/30-2\, b_3/15+b_4/20,\qquad\qquad\quad\qquad\ \,
 &b_{11}=-b_{10}=10\, b_9.
\end{eqnarray}  }
\indent
 First of all, as it is evident from
constraint~(\ref{cgnliii}), if any one of the $K_9$ or $K_{10}$ or
$K_{11}$ Lagrangian terms is missing at the beginning (i.e., any
one of the $b_9$ or $b_{10}$ or $b_{11}$ coefficients is zero),
then the trace analogy will hold only if all of these terms
vanish, and this case will return us to the case of part~(a). This
is confirmed once we set $b_9=0$ in constraint~(\ref{cgnliii}),
whereby it reduces to constraint~(\ref{cnliii}). Actually, we have
checked this point straightforward by omitting one of the $K_9$ to
$K_{11}$ terms just from the beginning. For example, by
considering only the $K_1$ to $K_{10}$ terms, and carrying out the
similar steps to what we have performed for the $K_1$ to $K_{11}$
terms, we find that the trace analogy imposes the constraint
 {\setlength\arraycolsep{10pt}
\begin{eqnarray}\label{cgnliv}
  b_3=0,\qquad\qquad\quad\
 &b_4=-2\, b_2/3,\qquad\qquad
 &b_5=-60\, b_1-8\, b_2,\cr
  b_6=30\, b_1+2\, b_2,\quad
 &b_7=30\, b_1+14\, b_2/3,\quad
 &b_8=100\, b_1+12\, b_2,\cr
  b_9=0=b_{10},\qquad\ \ &&
\end{eqnarray}  }
which satisfies constraint~(\ref{cgnliii}). Indeed, it is
constraint~(\ref{cnliii}) when its $b_3$ coefficient vanishes,
i.e. it is the corresponding case of~(\ref{cnliiis}) when the
$b_3$ coefficient is zero instead of the $b_8$ one.

Also, one can easily get the same result by considering the
outcomes of Ref.~\cite{Oliva2010} in where they have also derived
almost the dimensional dependent constraint version for the trace
relation (through a proportionality parameter, say $u$) via
classifying the six derivative Lagrangians of gravity whose the
traced field equations have a reduced order (see the Appendix~B).
They have employed the $K_1$ to $K_8$ terms plus $K_{15}$,
$K_{16}$, $K_{17}$ (that by identities (\ref{a8}), (\ref{a9}) and
(\ref{a10}) are equivalent to $K_{9}$, $K_{10}$ and $K_{11}$,
respectively) and $K_{13}$ terms, even though they have also
mentioned that these terms are~not all linearly independent
because of relations (\ref{a21}) and (\ref{a12}). Now, if one sets
the coefficients of the $K_{11}$ and $K_{13}$ terms simultaneously
zero in their result (their relation~(B15)), then the coefficients
of the $K_9$ and $K_{10}$ terms will vanish as well and the
remaining relations (after substituting for their proportionality
parameter in terms of the other coefficients, namely
$u=6(D-1)b_1+2b_2$\ ) will reduce to
constraint~(\ref{cnliiia})\rlap.\footnote{Note that, if one sets
only the coefficient of the $K_{13}$ term zero and $D=6$ in their
result, and also substitutes for their proportional parameter in
terms of the other coefficients, one will get
constraint~(\ref{cgnliii}).}

Yet, one may argue that identity~(\ref{a21}) can equally be used
for one of the other terms $K_4$ to $K_8$ as well. However, in
such a case, if one substitutes for, e.g. and without lost of
generality, the $K_8$ Lagrangian term, then the calculations will
give constraint~(\ref{cgnliii}) when its $b_8$ coefficient is
zero. Now, if one also sets $b_9=0$ in this new constraint, it
will reduce to constraint~(\ref{cnliiis}) which is an especial
case of part~(a) in six dimensions. Therefore, in order to have a
more general situation, it is more adequate to consider all the
$K_1$ to $K_{11}$ terms simultaneously\rlap.\footnote{The same
third order terms (though in different combinations) have also
been used in Ref.~\cite{kmm} (its relations $(2.12)$--$(2.17)$) as
all six--dimensional dimensionless actions.}

Secondly, the trace analogy for the introduced $G^{(3a)}_{({\rm
total})\alpha\beta}$ gives the number of independent $b_i$'s to be
four. This confirms an earlier analysis based on the cohomological
point of view (see relation~(4.13) of Ref.~\cite{bpb}) which also
gives consistency conditions (again see the invariance
condition~(4.10) of Ref.~\cite{bpb} for numerical linear
combinations of the first ten unknowns $b_i$). Also, the maximum
number of permissible missing coefficients of $b_1$ to $b_8$ is
three in constraint~(\ref{cgnliii}). However, once again in six
dimensions, due to relation~(\ref{eeight}), one also has another
extra constraint.

Let us now examine constraint~(\ref{cgnliii}) for the numerical
coefficients of a few available Lagrangians. As mentioned in
part~(a), Deser {\it et al.}~\cite{desch} (relation $(25c)$ of
their paper) and Karakhanyan {\it et al.}~\cite{kmm} (relation
$(2.18)$ of their paper) state that the corresponding Lagrangian
density of the expression (after adapting the sign convention)
\begin{equation}\label{edeser}
A_3\equiv C^{\mu\nu\rho\tau}\,\Square\, C_{\mu\nu\rho\tau}
    -2\, C^{\mu\nu\rho\alpha}C_{\mu\nu\rho\beta}R^{\beta}{}_\alpha
    +3\, C_{\mu\nu\rho\tau}R^{\mu\rho}R^{\nu\tau}
    +{3\over 2}\, K_4-{27\over 20}\, K_2+{21\over 100}\, K_1
\end{equation}
is also a Weyl--invariant combination in six
dimensions\rlap.\footnote{For the Weyl--invariant expressions in
arbitrary dimensions, see
Refs.~\cite{Oliva2010,Erdmenger97,ParkRose87}.}\
 After substituting for the Weyl tensor in six dimensions, it reads
\begin{equation}\label{edeserb}
A_3={41\over 100}\, K_1-{15\over 4}\, K_2+{9\over 2}\, K_4+5\,
    K_5-2\, K_6+{1\over 10}\, K_9-K_{10}+K_{11}.
\end{equation}
The coefficients of (\ref{edeserb}) satisfy
constraint~(\ref{cgnliii}). Another similar expression is given in
relation~(4.7) of Ref.~\cite{bpb} as (after adapting the sign
convention)
\begin{equation}\label{ebonora}
M_3\equiv -K_1+8\, K_2+2\, K_3-10\, K_4-10\, K_5-{1\over 2}\,
K_9+5\,
    K_{10}-5\, K_{11},
\end{equation}
that also satisfies constraint~(\ref{cgnliii}).

Karakhanyan {\it et al.}~\cite{kmm} also claim that there is an
additional\footnote{However, the relations $A_3$ and $A_4$
differ~\cite{kmm} from $A_1$ and $A_2$ in that they have non--zero
Weyl variations and one can employ them (only) as constraints on
local counterterms, but those cannot be considered as independent
contributions into the anomaly.}\
 Weyl--invariant action in six dimensions with the scalar
Lagrangian (relation $(2.19)$ of their paper, after adapting the
sign convention)
\begin{equation}\label{a4kmm}
A_4\equiv K_{10}-{3\over 10}\, K_9
    +2\, C_{\mu\nu\rho\tau}R^{\mu\rho}R^{\nu\tau}
    -K_4+{1\over 10}\, K_2+{1\over 50}\, K_1.
\end{equation}
Again, after substituting for the Weyl tensor in six dimensions,
it reads
\begin{equation}\label{a4kmmi}
A_4={3\over 25}\, K_1-K_2+2\, K_5-{3\over 10}\, K_9+K_{10}.
\end{equation}
Now, if one properly inserts the $K_{11}$ term from
identity~(\ref{a21}) into relation~(\ref{a4kmmi}), then it will
read effectively as\footnote{Indeed, relation (\ref{a4kmmi})
should be the effective one.}
\begin{equation}\label{a4kmmii}
A_4\mathrel{\mathop=^{\rm eff}}{3\over 25}\, K_1-K_2+{4\over 3}\,
K_4+{2\over 3}\, K_5-{2\over 3}\, K_6+{2\over 3}\, K_7+{4\over
3}\, K_8+{1\over 30}\, K_9-{1\over 3}\, K_{10}+{1\over 3}\,
K_{11},
\end{equation}
that its coefficients satisfy constraint~(\ref{cgnliii}).

Arakelyan {\it et al.}~\cite{akmm} give in relation~(18) of their
preprint, as the third linear cocycle, the expression (after
adapting the sign convention)
\begin{eqnarray}\label{earakel}
S^3_C\equiv\int\Bigl[\!\!\!\!\!\!\!\!\!\!
 &&C^{\alpha\beta\mu\nu}\,\Square\, C_{\alpha\beta\mu\nu}+4\,
   C^{\alpha\mu\nu\rho}C_{\beta\mu\nu\rho}R^{\beta}_\alpha-{6\over 5}
   C^{\alpha\mu\nu\rho}C_{\alpha\mu\nu\rho}R\cr
 &&+8\bigl(C^{\alpha\mu\nu\rho}C_{\beta\mu\nu\rho}\bigr)^{;\,\beta}{}_\alpha
   -{1\over 2}\Square\,\bigl(C^{\alpha\mu\nu\rho}C_{\alpha\mu\nu\rho}
   \bigr)\Bigr]\sqrt{-g}\, d^6\! x\equiv\int L^3_C\sqrt{-g}\, d^6\! x\, .
\end{eqnarray}
Substituting for the Weyl tensor in six dimensions, the used
Lagrangian, apart from the complete divergent terms, is actually
\begin{equation}\label{earakelb}
L^3_C\mathrel{\mathop=^{\rm eff}}-{11\over 50}\, K_1+{27\over
10}\, K_2-{6\over 5}\, K_3-3\, K_4-4\, K_5+4\, K_6+{1\over 10}\,
K_9-K_{10}+K_{11}\, ,
\end{equation}
that also satisfies constraint~(\ref{cgnliii}). Another similar
conformal anomaly is given in relation~(5) of Ref.~\cite{OdiRom94}
as (after adapting the sign convention and correcting a minor
mistype)
\begin{eqnarray}\label{odiromeo}
T_A\equiv\frac{1}{540\times
(4\pi)^3}\int\Bigl[\!\!\!\!\!\!\!\!\!\!
 &&-\frac{1}{300}K_1+\frac{1}{10}K_2-\frac{1}{10}K_3-\frac{3}{7}K_4+\frac{5}{21}K_5
   -\frac{2}{21}K_6+\frac{2}{7}K_7-\frac{23}{210}K_9\cr
 &&+\frac{13}{42}K_{10}+\frac{1}{12}K_{11}\Bigr]\sqrt{-g}\, d^6\! x\equiv
   \frac{1}{540\times (4\pi)^3}\int L_A\sqrt{-g}\, d^6\! x\, ,
\end{eqnarray}
where, by properly inserting the $K_{11}$ term from
identity~(\ref{a21}), the Lagrangian effectively reads
\begin{equation}\label{odiromeob}
L_A\mathrel{\mathop=^{\rm
eff}}-\frac{1}{300}K_1+\frac{1}{10}K_2-\frac{1}{10}K_3+\frac{2}{21}K_4-\frac{2}{7}
K_5-\frac{5}{14}K_6+\frac{23}{42}K_7+\frac{11}{21}K_8+\frac{3}{140}K_9
-\frac{3}{14}K_{10}+\frac{3}{14}K_{11}\, .
\end{equation}
These coefficients satisfy constraint~(\ref{cgnliii}).

Also, two Weyl--invariants (denoted by $\Sigma$ and $\Theta$) have
been introduced in arbitrary dimensions in Ref.~\cite{Oliva2010}
(their relations (B17) and (B18)), which in six dimensions are
\begin{eqnarray}\label{oliva}
\Theta{\bigr |}_{\rm 6-dim.}\!\!\!\!\!\!\!\!\!\!\!
 &&=-8\Sigma{\bigr |}_{\rm 6-dim.}\cr
 &&\equiv -8\left(-\frac{4}{25}K_1+2\, K_4+6\, K_5-8\, K_7-16\,
   K_8-\frac{9}{10}K_{15}+3\, K_{16}+{\rm total\
   derivative}\right)\!.
\end{eqnarray}
Using identities~(\ref{a8}) and~(\ref{a9}), and again properly
inserting the $K_{11}$ term from identity~(\ref{a21}), the
conformal anomaly~(\ref{oliva}) effectively reads
\begin{equation}\label{olivab}
\Sigma{\bigr |}_{\rm 6-dim.}\!\!\mathrel{\mathop=^{\rm
eff}}-\frac{4}{25}K_1-2\, K_4+10\, K_5+2\, K_6-10\, K_7-20\,
K_8-\frac{1}{10}K_9+K_{10}-K_{11},
\end{equation}
that its coefficients satisfy constraint~(\ref{cgnliii}).

In the next section, we examine the outcome of our rigorous
pursuit of the trace analogy approach for the leaded trace anomaly
(which also indulges into the heat kernel) from the third order
terms in six dimensions.

\section{Classical Trace Anomaly}
\indent

As mentioned in the Introduction, by enforcing the mathematical
appearance of the alternative form of the Einstein field
equations, i.e. $R_{\alpha\beta}=\kappa^2\, S_{\alpha\beta}/2$
where the source tensor is $S_{\alpha\beta}\equiv
T_{\alpha\beta}-g_{\alpha\beta}\,T/(D-2)$ in a $D$--dimensional
space--time with $T={\rm Trace}\,T_{\alpha\beta}$, for the
relevant alternative form of the Lovelock field equations, we have
classically justified~\cite{farc} that one gets
\begin{equation}\label{alove}
R^{\rm (Lovelock)}_{\alpha\beta}=\frac{1}{2}\kappa^2\, S^{\rm
(genera.)}_{\alpha\beta},
\end{equation}
where the generalized source tensor is
\begin{equation}\label{gsource}
S^{\rm (genera.)}_{\alpha\beta}\equiv
T_{\alpha\beta}-\frac{1}{D-2}g_{\alpha\beta}\left(T+T_{\rm
anomaly}\right)
\end{equation}
and
\begin{equation}\label{loveanomaly}
T_{\rm anomaly}=-\kappa^{-2}\,D\,\sum^{n_{_{\rm ext.}}}_{n\geq
1}\,{n-1\over n}\, c_n\, R^{(n)}\equiv \sum^{n_{_{\rm
ext.}}}_{n\geq 1}\, T^{(n)}_{\rm anomaly}\, .
\end{equation}
This shows that $T^{(1)}_{\rm anomaly}=0$, as expected to be for
the Einstein gravity, and gives the dimension of the $c_n$ to be
length to the power of $2(n-1)$, as indicated below the Lovelock
Lagrangian~(\ref{love}).

Then, we have applied~\cite{farc} the outcome for the generic
cases, i.e. $R^{(n)}_{\rm generic}$, with the appropriate
constraint condition on the coefficients and no upper limit for
$n$. Actually, we have examined only the resulted second order
trace anomaly in Ref.~\cite{farc}. That is
\begin{equation}\label{dttg}
T^{(2)}_{\rm anomaly}=-{\kappa^{-2}\,D\over 2}\, c_2\,
R^{(2)}_{\rm generic}\, ,
\end{equation}
where~\cite{farc,fare}
\begin{equation}\label{rgii}
R^{(2)}_{\rm generic}\equiv \kappa^2\,L^{(2)}_{\rm
generic}-\bigl(4\, a_1+a_2\bigr)\,\Square\, R=a_1 R^2+a_2
R_{\mu\nu}R^{\mu\nu}+a_3 R_{\alpha\beta\mu\nu}\,
R^{\alpha\beta\mu\nu}-\bigl(4\, a_1+a_2\bigr)\,\Square\, R\, ,
\end{equation}
with the constraint
\begin{equation}\label{cnlii}
3\, a_1+a_2+a_3=0
\end{equation}
that leaves two degrees of freedom in $D>4$ dimensions, or with
the constraint, e.g., $\alpha_2+\alpha_3=0$ with one degree of
freedom in (and up to) four dimensions (due to the Gauss--Bonnet
term), where $\alpha_2$ and $\alpha_3$ are just the new
coefficients for the corresponding terms.

Now in this work, we investigate the third order trace anomaly
resulted from the introduced total third order
Lagrangian~(\ref{lagtotal}). The trace anomaly issue for this case
is
\begin{equation}\label{dttt}
T^{(3)}_{\rm anomaly}=-{2\,\kappa^{-2}\, D\over 3}\, c_3\,
R^{(3a)}_{\rm total}\,,
\end{equation}
with constraint~(\ref{cgnliii}) and, as defined
in~(\ref{dgtotal}),
\begin{equation}\label{rthrt}
R^{(3a)}_{\rm total}\equiv \kappa^{2}\,L^{(3)}_{\rm
generic}+M^{(3)}+b_9\, R^{(K_9)}+b_{10}\, R^{(K_{10})}+b_{11}\,
R^{(K_{11})}.
\end{equation}
The explicit expression of~(\ref{rthrt}), after substituting
from~(\ref{liii}), (\ref{mlt}), (\ref{a15}), (\ref{a17})
and~(\ref{a18}) when considering constraint~(\ref{cgnliii}), is
\begin{eqnarray}\label{emehr}
R^{(3a)}_{\rm total}=\!\!\!\!\!\!\!\!
 &&b_1\, K_1+b_2\, K_2+b_3\,
   K_3-\left(120\, b_1+{244\over 15}b_2
   +{104\over 15}b_3+{12\over 5}b_4\right)K_4\cr
 &&+\left(60\, b_1+{122\over 15}b_2+{52\over 15}b_3+{16\over
   5}b_4\right)K_5-\left(8\,b_2+8\, b_3+6\, b_4\right)K_6\cr
 &&+\left(60\, b_1+{212\over 15}b_2+{142\over 15}b_3+{26\over
   5}b_4\right)K_7+\biggl(160\, b_1+{144\over 5}b_2+{104\over 5}b_3\cr
 &&+{36\over 5}b_4\biggr)K_8-\left(12\, b_1+b_2\right)K_9
   +\left(120\, b_1+{58\over 5}b_2+{48\over 5}b_3-{3\over 5}b_4\right)K_{10}\cr
 &&+\left(4\, b_2+3\, b_4\right)K_{11}-\left(60\, b_1+{152\over
   15}b_2+{52\over 15}b_3+{16\over 5}b_4\right)K_{12}\cr
 &&-\left(180\, b_1+{476\over 15}b_2+{316\over 15}b_3+{43\over
   5}b_4\right)K_{13}+\left({1\over 5}b_2-{4\over 5}b_3+{3\over 10}b_4\right)K_{14}\cr
 &&+\left(-12\, b_1-{61\over 30}b_2+{2\over 15}b_3-{4\over
   5}b_4\right)K_{15}+\biggl(180\, b_1+{137\over 5}b_2+{112\over 5}b_3\cr
 &&+{51\over 10}b_4\biggr)K_{16}+\left(-15\, b_1-{1\over 5}b_2-{11\over
   5}b_3+{6\over 5}b_4\right)K_{17}\, ,
\end{eqnarray}
with four degrees of freedom in $D>6$ dimensions. By considering
identities~(\ref{a11}), (\ref{a12}) and (\ref{a8})--(\ref{a10}),
it effectively reads (though, we have purposely kept the complete
divergent term~$K_{14}$)
\begin{eqnarray}\label{emehref}
R^{(3a)}_{\rm total}\mathrel{\mathop=^{\rm eff}}\!\!\!\!\!\!\!\!
 &&b_1\, K_1+b_2\, K_2+b_3\, K_3+\left(60\, b_1+{232\over 15}b_2
   +{212\over 15}b_3+{31\over 5}b_4\right)K_4\cr
 &&-\left(120\, b_1+{118\over 5}b_2+{88\over 5}b_3+{27\over
   5}b_4\right)K_5-\left(8\,b_2+8\, b_3+6\, b_4\right)K_6\cr
 &&+\left(60\, b_1+{212\over 15}b_2+{142\over 15}b_3+{26\over
   5}b_4\right)K_7+\biggl(160\, b_1+{144\over 5}b_2+{104\over 5}b_3
   +{36\over 5}b_4\biggr)K_8\cr
 &&+\left(15\, b_1+{39\over 10}b_2+{17\over
   5}b_3+{27\over 20}b_4\right)K_9-\left(60\, b_1+{79\over 5}b_2+{64\over
   5}b_3+{57\over 10}b_4\right)K_{10}\cr
 &&+\left(15\, b_1+{21\over 5}b_2+{11\over 5}b_3+{9\over 5}b_4\right)K_{11}
   +\left({1\over 5}b_2-{4\over 5}b_3+{3\over 10}b_4\right)K_{14}\, .
\end{eqnarray}

On the other hand, in the quantum aspects of gravity, it has been
shown that for $D=6(=2\times 3)$ dimensions, the finite and
renormalized expectation value of the trace of the
energy--momentum tensor, $\big\langle
T_\rho{}^\rho\big\rangle_{\rm ren}$, would have to be~\cite{dufc}
{\it cubic} in curvature, and so on for $D=2n$ dimensions to be to
the $n$th power in curvature\rlap.\footnote{For a brief review of
this subject see, e.g., Ref.~\cite{farc} and references therein.}\
 This effect is despite the
fact that the classical energy--momentum tensors, for the
conformally invariant classical actions, must be traceless. And,
it is known as a conformal, or trace, or Weyl anomaly, originally
noticed in 1973~\cite{capperduf}, which plays an important role in
understanding of many phenomena~\cite{dufc,deser96,NojOdin01}.
Actually, the anomalies generally occur in any regularization
method as a consequence of introducing a scale into the theory in
order to regularize it, see, e.g., Ref.~\cite{bida}. The
contribution of a divergent Lagrangian to the trace of the
energy--momentum tensor is one of the above consequences. Of
course, when the effective action is itself a conformally
invariant action, the expectation value of the trace of the {\it
total} energy--momentum tensor is zero.

We have indicated~\cite{farc} that constraint (\ref{cnlii}) is
exactly the same as the consistency condition on the numerical
coefficients that Duff suggested~\cite{dufa} in the process of
re--examining the Weyl anomaly applications when the dimensional
regularization is applied to a classically conformally invariant
theory in arbitrary dimension. Also, relation~(\ref{dttg}) gives
exactly the same result as in Ref.~\cite{ddichr}, in where it has
been shown that the relevant most general form of the anomalous
trace of the energy--momentum tensor for classically conformally
invariant fields of arbitrary spin and dimension is
\begin{equation}\label{fanom}
\big\langle T_\rho{}^\rho\big\rangle_{\rm ren}=-{\hbar\, c\over
180(4\pi)^2}\Bigl(a_1 R^2+a_2 R_{\mu\nu}R^{\mu\nu}+a_3
R_{\alpha\beta\mu\nu}\, R^{\alpha\beta\mu\nu}+\gamma\,\Square\,
R\Bigr).
\end{equation}
By comparison, it obviously shows that $\gamma=-\bigl(4\,
a_1+a_2\bigr)$, which completes the trace anomaly relations
suggested by Duff~\cite{dufa} and, in four dimensions, reveals
$c_2\propto\ell_P{}^2$, as expected.

The trace anomalies are~\cite{dufc} precisely the ${\bf b_m}$
coefficients ${\Bigl[}$also referred to as Hamidew (after
Hadamard--Minakshisundaram--DeWitt)~\cite{gibb} or, HMDS (after
the same persons plus Seely)~\cite{avra} or,
Minakshisundaram--Pleijel~\cite{aboc} coefficients${\Bigr]}$ of
the Schwinger--DeWitt proper time method, see, e.g.,
Ref.~\cite{avra} and references therein. These are the
t--independent terms in the asymptotic expansion of the heat
kernel\footnote{It is a very powerful tool in the mathematical
physics as well as in the quantum field theory.}\
 with the {\it appropriate} differential operator
$\triangle$, see, e.g., Refs.~\cite{bgzv,gilk}, in
\begin{equation}\label{appdila}
{\rm trace}\> e^{-\triangle t}\,\sim \sum^\infty_{\scriptstyle
m=0\atop\scriptstyle {\rm even\ no.}} B_m\, t^{m-D\over 2}\
\qquad\quad  t\to 0^+\, ,
\end{equation}
where $B_m=\int {\bf b_m}\,\sqrt{-g}\, d^D\! x$\ are invariants of
the differential operator and vanish for odd numbers of the $m$.
The calculation of the first order of these coefficients by the
pioneering method of DeWitt~\cite{DeWit65} is quite simple, but
gets very cumbersome at higher orders. Indeed, due to the
combinatorial explosion in the number of terms in the ${\bf b_m}$
and in the auxiliary tensorial quantities, improvement in the
higher orders $m$ has been tedious. Though, new algorithms and
computer algebra with improvements in computer systems have
appeared to perform great efficiency, see, e.g.,
Refs.~\cite{fkwc92,AvrSch95} and references therein.

As in the literature, there is~not, up to now, an explicit
calculated expression for the trace anomaly in six dimensions in
the semi--classical theory, we examine our third order trace
anomaly result~(\ref{dttt}) by a straightforward use of the ${\bf
b_6}$ coefficient of the Schwinger--DeWitt proper time method.
Though, let us first apply this comparison, as a re--examining,
for the second order one with the ${\bf b_4}$ coefficient.

For this purpose, if the $\triangle$ in (\ref{appdila}) being the
simplest of such an operator, e.g., the conformally invariant
Laplacian type operator\footnote{We have checked the signs with
Ref.~\cite{schim}, and $\xi(D)=(D-2)/[4(D-1)]$.}
\begin{equation}\label{lapope}
\triangle=\Square -\xi(D)\, R,
\end{equation}
then the ${\bf b_4}$, when the conformal coupling constant is
$\xi(4)=1/6$, will be given by\footnote{See, e.g., the relation
$E_4$ of Ref.~\cite{gilk} when its $E=R/6$ and $W_{ij}=0$, and
also adapting the sign convention.}
\begin{equation}\label{bfour}
{\bf b_4}=-{\hbar\, c\over 180(4\pi)^2}\left(R_{\mu\nu}R^{\mu\nu}
-R_{\alpha\beta\mu\nu}\, R^{\alpha\beta\mu\nu} -\Square\,
R\right),
\end{equation}
which is the trace anomaly in the case of massless conformal
scalar fields in four dimensions. On the other hand, the trace
anomaly~(\ref{dttg}), by substituting~(\ref{rgii}) with
constraint~(\ref{cnlii}), gives
\begin{equation}\label{dttgi}
T^{(2)}_{\rm anomaly}=-{\kappa^{-2}\,D\over 2}\,
c_2\left[-\frac{1}{3}\left(a_2+a_3\right)R^2+a_2
R_{\mu\nu}R^{\mu\nu}+a_3 R_{\alpha\beta\mu\nu}\,
R^{\alpha\beta\mu\nu}+\frac{1}{3}\left(a_2+4a_3\right)\,\Square\,
R\right],
\end{equation}
that, using the Gauss--Bonnet theorem~\cite{cheachebkonospiv} in
four dimensions, effectively reads
\begin{equation}\label{dttgii}
T^{(2)}_{\rm anomaly}\mathrel{\mathop=^{\rm eff}}-2\kappa^{-2}\,
c_2\left[-\frac{1}{3}\left(a_2+4a_3\right)\right]\left(R_{\mu\nu}R^{\mu\nu}
-R_{\alpha\beta\mu\nu}\, R^{\alpha\beta\mu\nu} -\Square\,
R\right).
\end{equation}
In comparison with~(\ref{bfour}), this consistently gives the
constraint $a_2+4a_3=-3$ (or, using~(\ref{cnlii}), equivalently
$4a_1+a_2=1$) with one degree of freedom and the same numerical
value for $c_2$, as expected.

Now, let us investigate the issue for the third order trace
anomaly. By using the relation $E_3$ of the theorem~$4.3$ of
Ref.~\cite{gilkb}, when its ${\cal E}=\xi(D)\, R$, the conformal
coupling constant is $\xi(6)=1/5$, the $W_{ij}=0$, a minor
mistyped is corrected\footnote{Its term $-4R_{ijik}R_{;\, jk}$
must read $-4R_{ijik}{\cal E}_{;\, jk}$, see also
Ref.~\cite{pato}.}\
 and adapting the sign convention, one gets
\begin{eqnarray}\label{egilkey}
{\bf b_6}\!=-{1\over 360(4\pi)^3}\Bigl(\!\!\!\!\!\!\!\!\!\!
  &&{1\over 450}K_1-{1\over
    15}K_2+{1\over 15}K_3-{4\over 63}K_4
    +{4\over 21}K_5+{8\over 21}K_6\cr
{}&&-{32\over 63}K_7-{40\over 63}K_8+{4\over 7}K_{10}-{6\over
    7}K_{11}-{2\over 35}K_{12}+{2\over 7}K_{13}\cr
{}&&-{3\over 35}K_{14}-{1\over 70}K_{15}+{1\over 7}K_{16}-{9\over
    14}K_{17}\Bigr).
\end{eqnarray}
This relation gives the trace anomaly in the case of massless
conformal scalar fields in six dimensions\rlap.\footnote{The
coefficient $-1/[360(4\pi)^3]$ is given in the natural units,
otherwise it reads $-\hbar^2\, G/[360(4\pi)^3\, c^2]$, for making
the dimension of the ${\bf b_6}$ coefficient to be the same as the
trace anomaly.}\
 The same issue has also been worked out by using an alternative method in
Ref.~\cite{aboc}, in where from their relation~$(3.3)$ with
$\xi(6)=1/5$ and using relation~(\ref{egk8}), one achieves
\begin{eqnarray}\label{eamst}
{\bf b_6}\!=-{1\over 360(4\pi)^3}\Bigl(\!\!\!\!\!\!\!\!\!\!
  &&{1\over 450}K_1-{1\over 15}K_2+{1\over 15}K_3-{4\over 63}K_4
    +{4\over 21}K_5-{4\over 3}K_6\cr
{}&&+{76\over 63}K_7+{176\over 63}K_8+{4\over 7}K_{10}-{2\over
    35}K_{12}+{2\over 7}K_{13}-{3\over 35}K_{14}\cr
{}&&-{1\over 70}K_{15}+{1\over 7}K_{16}-{9\over 14}K_{17}
    +{24\over 7}R_{\mu\nu;\,\lambda\rho}R^{\lambda\mu\nu\rho}
    \Bigr),
\end{eqnarray}
which is exactly the same as relation~(\ref{egilkey}) once the
last relation of~(\ref{useful11}) is substituted. However, by
considering identities~(\ref{a11}), (\ref{a12}) and
(\ref{a8})--(\ref{a10}), the trace anomaly relation
(\ref{egilkey}) effectively reads
\begin{eqnarray}\label{egilkeyi}
{\bf b_6}\!\mathrel{\mathop=^{\rm eff}}-{1\over
360(4\pi)^3}\Bigl(\!\!\!\!\!\!\!\!\!\!
  &&{1\over 450}K_1-{1\over
    15}K_2+{1\over 15}K_3-{22\over 63}K_4
    +{10\over 21}K_5+{8\over 21}K_6\cr
{}&&-{32\over 63}K_7-{40\over 63}K_8-{3\over 35}K_9+{3\over
7}K_{10}-{3\over
    14}K_{11}-{3\over 35}K_{14}\Bigr).
\end{eqnarray}

On the one hand, by setting the coefficients $b_1=1/450$,
$b_2=-1/15$, $b_3=1/15$ and\footnote{Note that, $b_1$, $b_2$ and
$b_3$ are evident from relation~(\ref{egilkeyi}) when are compared
with relation~(\ref{emehref}), and $b_4$ can be found from the
coefficient of the term~$K_6$ when one matches these relations in
the case of $b_2=-1/15=-b_3$.}\
 $b_4=-4/63$ in our relation~(\ref{emehref}), we also get in six dimensions
(though, we have~not used identity~(\ref{eeight}) yet),
\begin{eqnarray}\label{taftt}
T^{(3)}_{\rm anomaly}\mathrel{\mathop=^{\rm eff}}-4\,\kappa^{-2}\,
c_3\Bigl(\!\!\!\!\!\!\!\!\!
  &&{1\over 450}K_1-{1\over 15}K_2+{1\over 15}K_3-{22\over 63}K_4
    +{10\over 21}K_5+{8\over 21}K_6\cr
{}&&-{32\over 63}K_7-{40\over 63}K_8-{3\over 35}K_9+{3\over
    7}K_{10}-{3\over 14}K_{11}-{3\over 35}K_{14}\Bigr).
\end{eqnarray}
As it is evident, our result~(\ref{taftt}) is exactly the same as
relation~(\ref{egilkeyi}), and by comparison we also have
$c_3\propto\ell_P{}^4$, as expected.

\section{Conclusions}
\indent

In our previous works, we have shown that the analogy of the
Einstein tensor splitting into two parts with the trace relation
between them can be performed not only for each separate
(homogeneous) term of the Lovelock tensor, but also for the
(whole) Lovelock tensor as a complete Lagrangian (and indeed, for
any inhomogeneous Euler--Lagrange expression that can be spanned
linearly in terms of homogeneous tensors), via a generalized trace
operator, as well~\cite{farb}. For the second term of the Lovelock
tensor, we have discovered that it is~not only this term, treated
as an Euler--Lagrange expression of a special combination of the
second order Lagrangian terms, that possesses this analogy and
satisfies the trace relation, but also there are the
Euler--Lagrange expressions of the other generic cases of the
second order Lagrangian terms whose constant coefficients satisfy
a specific constraint, i.e. either exactly the Duff trace anomaly
relation or a dimensional dependent version of
it~\cite{farc,fare}. We have extended~\cite{farc} the analogy
further, and have manifested that the analogy of the alternative
form of the Einstein field equations for the relevant alternative
form of the Lovelock field equations reveals a classical approach
toward the trace anomaly with an indication of the constitution of
the higher order gravities towards it. Indeed, we have explicitly
shown~\cite{farc} that this procedure for any generic coefficients
of the second order term of the Lovelock Lagrangian yields exactly
the Duff trace anomaly relation, and even have
achieved~\cite{fare} a dimensional dependent version of this
relation.

In this work, we have probed further the analogy for the generic
coefficients of the eight terms in the third order of the Lovelock
Lagrangian, and have found the constraint relations between the
non--zero constituent coefficients into two forms, an independent
and a dimensional dependent versions. Each form has three degrees
of freedom, and the dimensional dependent constraint in six
dimensions is exactly the same as the other one. They do~not allow
simultaneously the missing of more than three coefficients. The
coefficients of the third order term of the Lovelock Lagrangian do
satisfy the two forms of the constraints, where in particular they
yield the dimensional dependent one in any dimension. The two
independent Lagrangian densities made from the cubic of the Weyl
tensor (as conformal invariants in six dimensions) also satisfy
the independent constraint only in six dimensions, and yield the
dimensional dependent version identically independent of the
dimension.

We have specified the all seventeen independent terms of the third
order type Lagrangian with the HDN three. Among these terms, we
have justified (by using a few complete divergent terms that lead
to the relevant identities) and have introduced the most general
effective expression as a total third order type Lagrangian with
arbitrary coefficients of just the eleven terms of them (the
previous eight terms plus the new three ones) denoted as the $K_1$
to $K_{11}$ in the text. Then, we have proceeded the analogy for
the field tensor of this combination, and have achieved the
relevant constraint among the non--zero constituent coefficients.
The constraint shows that, if one of these three new Lagrangian
terms is missed, then all of them will vanish (whereby the
constraint reduces to the previous case). Also, the maximum number
of permissible missing coefficients of the first previous eight
ones is again three. There are, in general, four degrees of
freedom, though in six dimensions, there exists another extra
identity among the first eight coefficients. We have shown that
the expressions given in the literature as the third
Weyl--invariant combination in six dimensions do satisfy the
obtained constraint relations. Thus, we suggest that these
constraint relations to be considered as the necessary consistency
conditions on the numerical coefficients that a Weyl--invariant
should satisfy (similar to the Duff consistency relations for the
second order trace anomaly).

We have reviewed the ``classical'' approach toward the trace
anomaly that was presented in our previous works, in order to
examine it for the introduced total third order type Lagrangian,
by then, we have derived the relevant trace anomaly. To examine
our outcome, as an explicit calculated expression for the trace
anomaly in six dimensions in the semi--classical theory has~not
been given in the literature, we have compared our result with the
precisely equivalent expression, namely the ${\bf b_6}$
coefficient of the Schwinger--DeWitt proper time method that
linked with the relevant heat kernel coefficient. For this
purpose, we first have achieved our general expressions for the
trace anomaly for the generic second order (as a re--examining
case) and then, for the total third order types Lagrangian terms,
relations~(\ref{dttgi}) and~(\ref{emehref}), with two and four
degrees of freedom (in $D>6$ dimensions, and three in six
dimensions), respectively. Then, we have demonstrated that the
obtained expressions contain exactly the ${\bf b_4}$ and ${\bf
b_6}$ coefficients, respectively, as a particular case. Of course,
these results are necessary consistency checks, nevertheless our
approach can be regarded as an alternative (perhaps simpler, and
classical) derivation of the trace anomaly which also gives a
general expression with the relevant degrees of freedom.

In aside, let us assert our view about the approach employed in
the work for general cases. The results obtained indicate that it
is likely that the analogy of the Einstein gravity should also
exist for a class of further generic Lagrangians of order/degree
$n>6$. However, for each order/degree to hold the analogy, there
would be a set of constraints that the relevant constituent
coefficients must satisfy. Of course, extending the analogy to
higher orders of generic cases does~not seem technically very easy
(even in the third order case considerable algebraic calculations
was required). That is, in order to find the constraints for
generic cases, or for the most general effective expressions as
total Lagrangian terms of each order, work must be performed order
by order with huge calculations for each one.

\section*{Acknowledgements}
\indent

We thank the Research Office of Shahid Beheshti University G.C.
for financial supporting of this work.
\setcounter{equation}{0}
\renewcommand{\theequation}{A.\arabic{equation}}
\section*{Appendix~A: Useful Relations \& Variation Of A Few Actions}
\indent

In this appendix, we furnish a few useful relations and also
supply the metric variation of a few Lagrangian terms.
\begin{itemize}
\item{\bf The Useful Relations}
\end{itemize}

The following derivative relations can easily be derived from
non--commutativity of the covariant derivatives, and the Bianchi
and the contracted Bianchi identities, as~\cite{fard}
\begin{eqnarray}\label{useful11}
{}&&R_{\alpha\mu}{}^{;\,\mu}{}_{\beta}=
    R_{;\,\alpha\beta}/2\, ,\cr
{}&&R_{\mu\alpha\nu\beta}{}^{;\,\alpha}
    =R_{\mu\nu ;\,\beta}-R_{\mu\beta ;\,\nu}\, ,\cr
{}&&R_{\mu\tau ;\,\nu}{}^{\tau}=R_{;\,\mu\nu}/2
    -R_{\mu\alpha\nu\beta}R^{\alpha\beta}+R_{\mu\alpha}
    R_{\nu}{}^{\alpha}\, ,\cr
{}&&R_{\nu\tau\mu\rho}R^{\tau\mu
    ;\,\nu\rho}=K_6/2-K_7/2-K_8-K_{11}/4
\end{eqnarray}
and also
\begin{equation}\label{useful22}
\bigl(R^{k-1}\bigr)_{;\,\mu\nu}=(k-1)R^{k-2}R_{;\,\mu\nu}
+(k-1)(k-2)R^{k-3}R_{;\,\mu}R_{;\,\nu}\, .
\end{equation}

In the variation process, using the integrating covariantly by
parts and the appropriate boundary conditions, one can effectively
write
\begin{equation}\label{useful33}
f\,\delta\bigl(\Square\,^k R\bigr)\mathrel{\mathop=^{\rm
eff}}\bigl(\Square\,^k f\bigr)\, \delta R
  +\sum_{i=0}^{k-1}\bigl(\Square\,^i f\bigr)\left[-\left(\Square\,^{(k-1-i)}
  R\right)_{,\,\tau}\,g^{\alpha\rho}\,\delta\,\Gamma^\tau{}_{\alpha\rho}
  +\left(\Square\,^{(k-1-i)} R\right)_{;\,\alpha\rho}\,\delta g^{\alpha\rho}
  \right],
\end{equation}
where the $f$ is any scalar function of the metric and its
derivatives.

The Euler--Lagrange expressions of a few scalar Lagrangian density
terms\rlap,\footnote{The corresponding Euler--Lagrange expressions
of Lagrangians containing the derivatives of the curvature scalar
are firstly due to Buchdahl~\cite{buchd}.}\
 where ${\cal L}=F\sqrt{-g}/\kappa^2$,
are~\cite{fard}
\begin{equation}\label{useful44}
G^{\bigl(F(R)\bigr)}_{\alpha\beta}=F^{\,\prime}{}\,
R_{\alpha\beta}
             -F^{\,\prime}{}_{;\,\alpha\beta}-{1\over 2}\, g_{\alpha\beta}
             \Bigl(F-2\,\Square\, F^{\,\prime}\Bigr),
\end{equation}
\begin{eqnarray}\label{useful55}
G^{\bigl(F(R_{\mu\nu}\,
R^{\mu\nu})\bigr)}_{\alpha\beta}\!=\!\!\!\!\!\!\!\!
  &&2F^{\,\prime}\, R_{\alpha\rho}\, R_{\beta}{}^\rho
    -\bigl(F^{\,\prime}\, R_{\alpha\rho}\bigr)_{;\,\beta}{}^\rho
    -\bigl(F^{\,\prime}\, R_{\beta\rho}\bigr)_{;\,\alpha}{}^\rho
    +\Square\, \bigl(F^{\,\prime}\, R_{\alpha\beta}\bigr)\cr
{}&&-{1\over 2}\, g_{\alpha\beta}\Bigl[F
    -2\bigl(F^{\,\prime}\, R^{\mu\nu}\bigr)_{;\,\mu\nu}\Bigr],
\end{eqnarray}
\begin{equation}\label{useful66}
G^{\bigl(F(R_{\rho\tau\mu\nu}\,
R^{\rho\tau\mu\nu})\bigr)}_{\alpha\beta}\!
  =2F^{\,\prime}\, R_{\alpha\rho\mu\nu}R_{\beta}{}^{\rho\mu\nu}
   +2\bigl(F^{\,\prime}\, R_{\mu\alpha\nu\beta}\bigr)^{;\,\mu\nu}
   +2\bigl(F^{\,\prime}\, R_{\mu\beta\nu\alpha}\bigr)^{;\,\mu\nu}
  -{1\over 2}\, g_{\alpha\beta}\, F\, ,
\end{equation}
\begin{eqnarray}\label{useful77}
G^{\bigl(F(\sum\limits^p_{k=0} \Square\,^k
R)\bigr)}_{\alpha\beta}\!=\!\!\!\!\!\!\!\!
  &&\Bigl(\sum_{k=0}^p\Square\,^k f_k\Bigr)R_{\alpha\beta}
    -\Bigl(\sum_{k=0}^p\Square\,^k f_k\Bigr)_{;\,\alpha\beta}
    -\sum_{k=1}^p\sum_{i=0}^{k-1}\bigl(\Square\,^i f_k\bigr)_{(;\,\alpha}
    \bigl(\Square\,^{(k-1-i)}R\bigr)_{;\,\beta)}\cr
{}&&-{1\over 2}\, g_{\alpha\beta}
    \biggl\{F-2\,\Square\Bigl(\sum_{k=0}^p\Square\,^k f_k\Bigr)
    -\sum_{k=1}^p\sum_{i=0}^{k-1}\Bigl[\bigl(\Square\,^i f_k\bigr)
    \bigl(\Square\,^{(k-1-i)}R\bigr)_{;\rho}\Bigr]^{;\rho}\biggr\}
\end{eqnarray}
and
\begin{eqnarray}\label{useful88}
G^{\bigl(R^N\,\Square\,^l
R\bigr)}_{\alpha\beta}\!=\!\!\!\!\!\!\!\!\!
  &&\Theta^{[Nl]}R_{\alpha\beta}-\Theta^{[Nl]}{}_{;\,\alpha\beta}
    -\sum_{i=0}^{l-1}\bigl(\Square\,^i R^N\bigr)_{(;\,\alpha}
    \bigl(\Square\,^{(l-1-i)}R\bigr)_{;\,\beta)}\cr
{}&&-{1\over 2}\, g_{\alpha\beta}\biggl\{R^N\,\Square\,^l
    R-2\,\Square\,\Theta^{[Nl]}-\sum_{i=0}^{l-1}\Bigl[\bigl(\Square\,^i R^N\bigr)
    \bigl(\Square\,^{(l-1-i)}R\bigr)_{;\rho}\Bigr]^{;\rho}\biggr\},
\end{eqnarray}
where the prime denotes ordinary derivative with respect to the
argument, $f_k\equiv
\partial F/\partial\,\Square\,^k R$ and $\Theta^{[Nl]}\equiv N\,R^{N-1}\,\Square\,^l R+\Square\,^l R^N$.

The Weyl conformal tensor, for $D\geq 3$ dimensions, satisfies
\begin{equation}\label{wsd}
C_{\alpha\beta\mu\nu}\,C^{\alpha\beta\mu\nu}=
{2\over(D-1)(D-2)}R^2-{4\over(D-2)}R_{\mu\nu}R^{\mu\nu}+
R_{\alpha\beta\mu\nu}\,R^{\alpha\beta\mu\nu},
\end{equation}
where in four dimensions
$C_{\alpha\beta\mu\nu}C^{\alpha\beta\mu\nu}{\Bigr |}_{\rm
4-dim.}\!\!=R^2/3-2R_{\mu\nu}R^{\mu\nu}+R_{\alpha\beta\mu\nu}\,
R^{\alpha\beta\mu\nu}$. And, its associated effective Lagrangian,
by considering the Gauss--Bonnet theorem, is
$-2/3\left(R^2-3R_{\mu\nu}R^{\mu\nu}\right)$. In six dimensions,
it reads $C_{\alpha\beta\mu\nu}C^{\alpha\beta\mu\nu}{\Bigr |}_{\rm
6-dim.}\!\!=R^2/10-R_{\mu\nu}R^{\mu\nu}+R_{\alpha\beta\mu\nu}\,
R^{\alpha\beta\mu\nu}$, also $C_{\alpha\beta\mu\nu}\Square\,
C^{\alpha\beta\mu\nu}{\Bigr |}_{\rm
6-dim.}\!\!=K_9/10-K_{10}+K_{11}$\ and $C_{\alpha\beta\mu\nu\,
;\rho}C^{\alpha\beta\mu\nu\, ;\rho}{\Bigr |}_{\rm
6-dim.}\!\!=K_{15}/10-K_{16}+K_{17}$. In three dimensions, as the
Weyl tensor identically vanishes, from relations~(\ref{cwtd})
and~(\ref{cwtda}) we have, respectively,
\begin{equation}\label{weylcub3d1}
\left(6 K_1-36 K_2+3 K_3+32 K_4+24 K_5-12 K_6+K_7\right){\Bigr
|}_{\rm 3-dim.}\!\!\equiv 0
\end{equation}
and
\begin{equation}\label{weylcub3d2}
\left(-7 K_1/2+21 K_2-3 K_3/2-18 K_4-15 K_5+6 K_6+K_8\right){\Bigr
|}_{\rm 3-dim.}\!\!\equiv 0.
\end{equation}
In four dimensions, from relations~(\ref{cwtd}) and~(\ref{cwtda})
and their dependence, one has
\begin{equation}\label{weylcub4d}
\left(\frac{8}{9}K_1+18 K_2-3 K_3-20 K_4-24 K_5+18 K_6-K_7+4
K_8\right){\Bigr |}_{\rm 4-dim.}\!\!\equiv 0.
\end{equation}
Also, see Ref.~\cite{Erdmenger97} and the appendix of
Ref.~\cite{bgzv}. The scalar action $I$ constructed by the cubic
of the Weyl tensor, relations~(\ref{cwtd}) or~(\ref{cwtda}), in a
$D$--dimensional space--time, through the conformal transformation
$g_{\mu\nu}\to\Omega^2g_{\mu\nu}$, conformally transforms as $I\to
\Omega^{D-6}\,I$. Hence, it is a conformal invariant only in six
dimensions.

\begin{itemize}
\item{\bf The $M^{(3)}$ As A Lagrangian Term}
\end{itemize}

In Sect.~$4$, we need to know the effect of the $M^{(3)}$,
relation (\ref{mlt}), as a Lagrangian term. By using the relations
\begin{equation}\label{a2}
\Square\, R^2=2\bigl(K_9+K_{15}\bigr),
\end{equation}
\begin{equation}\label{a3}
\Square\,\bigl(R_{\mu\nu}R^{\mu\nu}\bigr)=2\bigl(K_{10}+K_{16}\bigr),
\end{equation}
\begin{equation}\label{a4}
\Square\,\bigl(R_{\mu\nu\rho\tau}R^{\mu\nu\rho\tau}\bigr)=2\bigl(K_{11}
+K_{17}\bigr),
\end{equation}
\begin{equation}\label{a5}
\Bigl(R^{\mu\nu}R_{;\,\mu}\Bigr)_{;\,\nu}=K_{12}+{1\over 2}K_{15}
\, ,
\end{equation}
\begin{equation}\label{a6}
\Bigl(R_{\mu\nu}R^{\mu\rho}\Bigr)_{;\,\rho}{}^\nu=
K_4-K_5+K_{12}+K_{13}+{1\over 4}K_{15}
\end{equation}
and
\begin{equation}\label{a7}
\Bigl(R_{\mu\nu;\,\rho}R^{\rho\mu\nu\lambda}\Bigr)_{;\,\lambda}={1\over
2}K_6-{1\over 2}K_7-K_8-{1\over 4}K_{11}+K_{13}-K_{16}\, ,
\end{equation}
relation (\ref{mlt}) reads as
\begin{eqnarray}\label{a1}
M^{(3)}\!=\!\!\!\!\!\!\!\!\!
   &&-\!{1\over 2}\biggl[\!\bigl(12b_1\!+\!b_2\bigr)R^2\!
     +\!2\bigl(2b_2\!+\!b_5\bigr)R_{\mu\nu}R^{\mu\nu}\!
     +\!\bigl(4b_3\!+\!{1\over 2}b_6\bigr)R_{\mu\nu\lambda\tau}
     R^{\mu\nu\lambda\tau}\biggr]_{;\rho}{}^\rho\cr
 {}&&-\bigl(2b_2+b_5\bigr)\Bigl(R^{\mu\nu}R_{;\,\mu}\Bigr)_{;\,\nu}
     -\bigl(3b_4-2b_5\bigr)\Bigl(R_{\mu\nu}R^{\mu\rho}\Bigr)_{;\,\rho}{}^\nu
     +2\bigl(b_5+b_6\bigr)\Bigl(R_{\mu\nu;\,\rho}R^{\rho\mu\nu\lambda}
     \Bigr)_{;\,\lambda}\,.
\end{eqnarray}
Obviously all of the terms are complete divergences. Therefore,
with careful attention about the appropriate boundary conditions
(that need to be applied in the process of the variation when a
function is up to the fourth order jet--prolongation of the
metric), one can easily show that these terms, and hence the
$M^{(3)}$, give no contribution to the variation of the
corresponding action.

\begin{itemize}
\item{\bf The $K_9$ To $K_{17}$ As Lagrangian Terms}
\end{itemize}

In Sect.~$5$, the effect of each linearly independent term of the
$K_9$ up to $K_{17}$, as a third order Lagrangian term, is needed.
For this purpose, using the null effect of complete
divergences~(\ref{a2})--(\ref{a7}) in the variation of their
corresponding action and that, the term $K_{14}$ is itself a
complete divergence, we obtain\footnote{Here, one actually means
that the identity holds between the functional derivatives, i.e.,
for example, $\delta(K_{16}\sqrt{-g})/\delta g^{\alpha\beta}\equiv
-\delta(K_{10}\sqrt{-g})/\delta g^{\alpha\beta}$.}\
 the following identities, i.e.
\begin{equation}\label{a21}
G^{(K_{11})}_{\alpha\beta}\equiv -4\, G^{(K_4)}_{\alpha\beta} +4\,
G^{(K_5)}_{\alpha\beta}+2\, G^{(K_6)}_{\alpha\beta} -2\,
G^{(K_7)}_{\alpha\beta}-4\, G^{(K_8)}_{\alpha\beta}
-G^{(K_9)}_{\alpha\beta}+4\, G^{(K_{10})}_{\alpha\beta}\, ,
\end{equation}
\begin{equation}\label{a11}
G^{(K_{12})}_{\alpha\beta}\equiv {1\over
2}G^{(K_9)}_{\alpha\beta}\, ,
\end{equation}
\begin{equation}\label{a12}
G^{(K_{13})}_{\alpha\beta}\equiv -G^{(K_4)}_{\alpha\beta}
+G^{(K_5)}_{\alpha\beta}-{1\over 4}G^{(K_9)}_{\alpha\beta}\, ,
\end{equation}
\begin{equation}\label{k14}
G^{(K_{14})}_{\alpha\beta}\equiv 0\, ,
\end{equation}
\begin{equation}\label{a8}
G^{(K_{15})}_{\alpha\beta}\equiv -G^{(K_9)}_{\alpha\beta}\, ,
\end{equation}
\begin{equation}\label{a9}
G^{(K_{16})}_{\alpha\beta}\equiv -G^{(K_{10})}_{\alpha\beta}
\end{equation}
and
\begin{equation}\label{a10}
G^{(K_{17})}_{\alpha\beta}\equiv -G^{(K_{11})}_{\alpha\beta}\, .
\end{equation}
Relation (\ref{a21}) has already been used in Ref.~\cite{gilk}
(the last relation on its page number 612, after adapting the sign
convention and substituting the necessary relations) and
Ref.~\cite{Oliva2010} (the first relation of their relation~(14)).
However, consideration caution has to be exercised in applying
identities in general, and identity~(\ref{a21}) in particular, for
there are associated with them some difficulties, but nonetheless
they throw some important effects on the effective actions.

Thus, it appears that among the $K_9$ to $K_{17}$ Lagrangian
terms, it suffices to derive only the metric variation of the
$K_9$ and $K_{10}$ terms; indeed, we mean that
\begin{equation}\label{a13}
\delta\int\sum^{17}_{i=1}b_i\, K_i\sqrt{-g}\, d^D\! x
=\delta\int\sum^{10}_{i=1}b^{'}_i\, K_i\sqrt{-g}\, d^D\! x\, ,
\end{equation}
where
 {\setlength\arraycolsep{3pt}
\begin{eqnarray}\label{a14}
 b^{'}_1=b_1,\qquad\quad\qquad\qquad\qquad\ \ \!\!
  &b^{'}_2=b_2,\qquad\quad\qquad\qquad\qquad
  &b^{'}_3=b_3,\quad\cr
 b^{'}_4=b_4-4b_{11}-b_{13}+4b_{17},\quad\ \!\!
  &b^{'}_5=b_5+4b_{11}+b_{13}-4b_{17},\ \,
  &b^{'}_6=b_6+2b_{11}-2b_{17},\cr
 b^{'}_7=b_7-2b_{11}+2b_{17},\qquad\qquad\!\!\!
  &b^{'}_8=b_8-4b_{11}+4b_{17},\qquad\ \ \,
  &b^{'}_9=b_9-b_{11}+b_{12}/2-b_{13}/4-b_{15}+b_{17},\cr
 b^{'}_{10}=b_{10}+4b_{11}-b_{16}-4b_{17}\, .&
\end{eqnarray}  }
Hence, as it is customary to write a Lagrangian in terms of its
effective one, one can write
\begin{equation}\label{leff}
\sum^{17}_{i=1}b_i\, K_i\mathrel{\mathop=^{\rm
eff}}\sum^{10}_{i=1}b^{'}_i\, K_i\, ,
\end{equation}
where also the word ``effective'' is often not mentioned. However,
as explained in the text, for being able to apply the trace
analogy for an effective total third order Lagrangian, we, in
practice, need to consider the $K_1$ to $K_{11}$ terms, and
actually neglect identity~(\ref{a21}).

The $K_9$, $K_{10}$ and $K_{11}$ Lagrangian terms give
\begin{equation}\label{a15}
G^{(K_9)}_{\alpha\beta}=\left[2\bigl(\Square\, R\bigr)
R_{\alpha\beta}-2\bigl(\Square\, R\bigr)_{;\,\alpha\beta}
-R_{;\,\alpha}R_{;\,\beta}\right]-{1\over 2} \,
g_{\alpha\beta}\Bigl(-4\, K_{14}-K_{15}\Bigr),
\end{equation}
\begin{equation}\label{a17}
G^{(K_{10})}_{\alpha\beta}=R^{(K_{10})}_{\alpha\beta}-{1\over 2}
\, g_{\alpha\beta}\Bigl(-2\, K_6+2\, K_7+4\, K_8+2\, K_{10}+K_{11}
-4\, K_{13}-K_{14}+5\, K_{16}\Bigr)
\end{equation}
and
\begin{equation}\label{a18}
G^{(K_{11})}_{\alpha\beta}=R^{(K_{11})}_{\alpha\beta}-{1\over 2}
  \, g_{\alpha\beta}\bigl(-K_{17}\bigr),
\end{equation}
where\footnote{As we will~not need the exact expressions of the
$R^{(K_{10})}_{\alpha\beta}$ and $R^{(K_{11})}_{\alpha\beta}$, to
reduce the amount of calculations we have derived their traces
almost from the beginning. Implicitly, the appearances of the
Euler--Lagrange expressions (\ref{useful44})--(\ref{useful88}) and
(\ref{a15})--(\ref{a18}) have been written according to part~(a).}
\begin{equation}\label{a16}
{\rm Trace}\, R^{(K_9)}_{\alpha\beta}={1\over 3}\Bigl(2\, K_9-2\,
K_{14}-K_{15}\Bigr),
\end{equation}
\begin{equation}\label{a19}
{\rm Trace}\, R^{(K_{10})}_{\alpha\beta}={1\over 3}\Bigl( 2\,
K_4-2\, K_5-2\, K_6+2\, K_7+4\, K_8+4\, K_{10}+K_{11} -2\,
K_{13}-{1\over 2}\, K_{15}+5\, K_{16}\Bigr)
\end{equation}
and
\begin{equation}\label{a20}
{\rm Trace}\, R^{(K_{11})}_{\alpha\beta}={1\over 3}\Bigl(8\,
K_6-8\, K_7 -16\, K_8-4\, K_{10}+16\, K_{13}+2\, K_{14}-20\,
K_{16}+K_{17}\Bigr).
\end{equation}
\setcounter{equation}{0}
\renewcommand{\theequation}{B.\arabic{equation}}
\section*{Appendix~B: Dimensional Dependent Constraints For $L^{(3)}_{\rm total}$}
\indent

In this appendix, we perform the analogy for the appearance of the
$G^{(3b)}_{({\rm total})\alpha\beta}$ in accord with part~(b),
where
\begin{equation}\label{dgtotalb}
G^{(3b)}_{({\rm total})\alpha\beta}=G^{(3b)}_{({\rm
generic})\alpha\beta}+b_9\, G^{(K_9)}_{\alpha\beta} +b_{10}\,
G^{(K_{10})}_{\alpha\beta}+b_{11}\, G^{(K_{11})}_{\alpha\beta}
\end{equation}
and similar expressions for $R^{(3b)}_{({\rm total})\alpha\beta}$
and $R^{(3b)}_{\rm total}$. The $G^{(3b)}_{({\rm
generic})\alpha\beta}$ is given by relation~(\ref{nliiia}) and the
Euler--Lagrange expressions of the~$K_9$ to $K_{11}$ Lagrangian
terms (that are given by relations~(\ref{a15})--(\ref{a18})) must
be rewritten in accord with part~(b). By imposing the condition
${\sl Trace}\, R^{(3b)}_{({\rm total})\alpha\beta}=R^{(3b)}_{\rm
total}$, and performing the bulk calculations, we find that the
trace relation, for non--zero coefficients, holds if and only if
\begin{eqnarray}\label{cnliiiab}
 &&b_5=\left[-12(D-1)^2b_1-(3D^2+3D-4)b_2/3-4(3D-5)b_3/3-b_4\right]/(D-1),\cr
 &&b_6=\left[6(D-1)^2b_1-(3D^2-19D+14)b_2/6-2(D+1)b_3/3-(3D^2-7D+2)b_4/4\right]/(D-1),\cr
 &&b_7=(D-2)\Bigl[3(D-1)^2b_1+(5D^2+3D-6)b_2/12+(3D^2-7D+6)b_3/3\cr
 &&\qquad\qquad\qquad +(D^2+3D-2)b_4/8\Bigr]/[2(D-1)],\cr
 &&b_8=4(D-1)^2b_1+(D-1)(D+2)b_2/3+4(D-1)b_3/3+(D-2)b_4/2,\cr
 &&b_9=\left(b_2/6-2b_3/3+b_4/4\right)/(D-1),\cr
 &&b_{10}=-2(D-1)b_9,\cr
 &&b_{11}=(D-1)(D-2)b_9/2=-(D-2)b_{10}/4.
\end{eqnarray}
If one sets one of the $b_9$ or $b_{10}$ or $b_{11}$ zero, then
constraint~(\ref{cnliiiab}) will reduce to
constraint~(\ref{cnliiia}). As mentioned in the text, the authors
of Ref.~\cite{Oliva2010} have also derived almost the same
relations through a proportionality parameter, say $u$. Indeed, if
we set the $b_{13}=0$ and substitute for their proportional
parameter in terms of the other coefficients, namely
$u=[6(D-1)^2b_1-(D^2-8D+6)b_2/3+4D(D-2)b_3/3-D(D-2)b_4/2]/(D-1)$,
in their relation (B15), it will reduce to
constraint~(\ref{cnliiiab}).

%
\end{document}